\documentclass[amsmath,10pt,twocolumn,superscriptaddress,footinbib,pre]{revtex4-1}
\usepackage[utf8]{inputenc}
\usepackage{hyperref}
\usepackage{graphicx}
\usepackage{tabularx}
\usepackage{amsmath,amsfonts,bm, amssymb}
\usepackage{uri}
\usepackage[version=4]{mhchem}

\usepackage[x11names, rgb, html]{xcolor}
\definecolor{sbase03}{HTML}{002B36}
\definecolor{sbase02}{HTML}{073642}
\definecolor{sbase01}{HTML}{586E75}
\definecolor{sbase00}{HTML}{657B83}
\definecolor{sbase0}{HTML}{839496}
\definecolor{sbase1}{HTML}{93A1A1}
\definecolor{sbase2}{HTML}{EEE8D5}
\definecolor{sbase3}{HTML}{FDF6E3}
\definecolor{syellow}{HTML}{B58900}
\definecolor{sorange}{HTML}{CB4B16}
\definecolor{sred}{HTML}{DC322F}
\definecolor{smagenta}{HTML}{D33682}
\definecolor{sviolet}{HTML}{6C71C4}
\definecolor{sblue}{HTML}{268BD2}
\definecolor{scyan}{HTML}{2AA198}
\definecolor{sgreen}{HTML}{859900}
\hypersetup{
  colorlinks = true,
  citecolor = {sred},
  urlcolor = {sblue}
}
\usepackage{oplotsymbl}
\usepackage{marvosym}
\usepackage{wasysym}
\usepackage{fdsymbol}
\usepackage{stmaryrd}

\begin{document}
\title{Limits on the Precision of Catenane Molecular Motors: Insights from Thermodynamics and Molecular Dynamics Simulations}
\author{Alex Albaugh}
\thanks{These authors contributed equally to this work.}
\affiliation{Department of Chemical Engineering and Materials Science, Wayne State University, 5050 Anthony Wayne Drive, Detroit, Michigan 48202, USA}
\author{Rueih-Sheng Fu}
\thanks{These authors contributed equally to this work.}
\affiliation{Department of Chemistry, Northwestern University, 2145 Sheridan Road, Evanston, Illinois 60208, USA}
\author{Geyao Gu}
\affiliation{Department of Chemistry, Northwestern University, 2145 Sheridan Road, Evanston, Illinois 60208, USA}
\author{Todd R.~Gingrich}
\email{todd.gingrich@northwestern.edu}
\affiliation{Department of Chemistry, Northwestern University, 2145 Sheridan Road, Evanston, Illinois 60208, USA}

\begin{abstract}
  Thermodynamic uncertainty relations (TURs) relate precision to the dissipation rate, yet the inequalities can be far from saturation.
  Indeed, in catenane molecular motor simulations, we record precision far below the TUR limit.
  We further show that this inefficiency can be anticipated by four physical parameters: the thermodynamic driving force, fuel decomposition rate, coupling between fuel decomposition and motor motion, and rate of undriven motor motion.
  The physical insights might assist in designing molecular motors in the future.
\end{abstract}

\maketitle

\section*{Introduction}
Molecular motors generate directed motion, extracting free energy from their environment and producing entropy in the process~\cite{brown2019theory}.
Biological motors like myosin~\cite{finer1994single}, dynein~\cite{schnapp1989dynein}, and ATP synthase~\cite{yasuda1998f1} are responsible for important processes like muscle contraction, molecular transport, and chemical fuel generation, respectively.
Recent breakthroughs in synthetic chemistry have also led to autonomous artificial motors that are chemically fueled~\cite{wilson2016autonomous,borsley2021doubly,borsley2022autonomous,korosec2021substrate,erbas2015artificial,unksov2022through,amano2021catalysis}.
With net flows of energy, molecular motor systems are necessarily out of equilibrium.
Systems driven only weakly out of equilibrium can be analyzed with linear response theory, but molecular motors do not necessarily operate in such a regime.
Limited tools are available for studying systems far from equilibrium, and the recent development of fluctuation theorems \cite{evans2002fluctuation, rao2018detailed, esposito2010three} and associated results have allowed for novel thermodynamic analyses of these systems \cite{seifert2005fluctuation, andrieux2006fluctuation, pietzonka2014fine}.

A family of such results, known collectively as thermodynamic uncertainty relations (TURs), governs the relationship between fluctuations in a time-extensive current \(J\) and the total dissipation \(\Sigma\).
TURs were first studied, postulated, and derived in the context of Markov jump processes in the long-time limit \cite{barato2015thermodynamic, gingrich2016dissipation, pietzonka2016universal} and have since been generalized \cite{falasco2020unifying} to a wide variety of domains, such as Markov chains~\cite{proesmans2017discrete}, diffusions~\cite{polettini2016tightening, gingrich2017inferring, van2019uncertainty, lee2021universal, fischer2018large}, and quantum systems~\cite{carollo2019unraveling, guarnieri2019thermodynamics, hasegawa2021thermodynamic}.
The classical overdamped TUR can be expressed in the form
\begin{equation}
\frac{\mathrm{var}(J)}{\langle J\rangle^2} \geq \frac{2 k_\mathrm{B}}{\langle\Sigma\rangle},
\label{eq:tur}
\end{equation}
where \(k_\mathrm{B}\) is the Boltzmann constant, \(\langle\cdot\rangle\) is the mean, and \(\text{var}(\cdot)\) is the variance.
A common biochemical situation is that the dissipation comes from the net decomposition of \(N_{\rm rxn}\) fuel molecules, each feeling a thermodynamic driving force \(\Delta \mu\) at temperature \(T\).
In that no-load case, the dissipation associated with the net reactions \(\Sigma_{\rm rxn} = N_{\rm rxn} \Delta \mu / T\) can be viewed as a particular current of interest, allowing Eq.~\eqref{eq:tur} to be translated into a restriction on the stochastic fluctuations in the number of fuel decomposition events: 
\begin{equation}
  \mathrm{var}(N_{\rm rxn}) \ge \frac{2}{\beta \Delta \mu} \langle N_{\rm rxn} \rangle,
\label{eq:tur_ent}
\end{equation}
where \(\beta = 1 / k_{\rm B} T\).
TURs set a fundamental limit on the precision of fluctuating systems, so it is natural to characterize a motor's efficiency by how closely it saturates the corresponding TUR \cite{song2021thermodynamic}.
This measure of efficiency is meaningful even when a motor spins with no load.
In that case, the thermodynamic efficiency, measured as work out per energy input, necessarily vanishes simply because there is no work with no load.
By contrast, the efficiency we discuss measures how effectively the motor generates directed motion, something that can occur even in the absence of the load.
Much of the literature surrounding TURs deals with relatively low-dimensional models and systems.
Here, we demonstrate how a high-dimensional particle-based model of an artificial molecular motor can be used in conjunction with molecular dynamics simulations to generate a direct comparison against the TUR and explain how it can be employed in studying molecular motors.
This effort complements current research on the optimal control of and performance trade-offs for molecular motors \cite{pietzonka2016universal2, leighton2022performance, gupta2022optimal, lathouwers2020nonequilibrium}.

\begin{figure*}
\centering
\begin{tikzpicture}
    \node[anchor=south west,inner sep=0] (image) at (0,0) {
      \includegraphics[trim=0.5cm -0.45cm 0.05cm 0,height=0.255\textwidth]{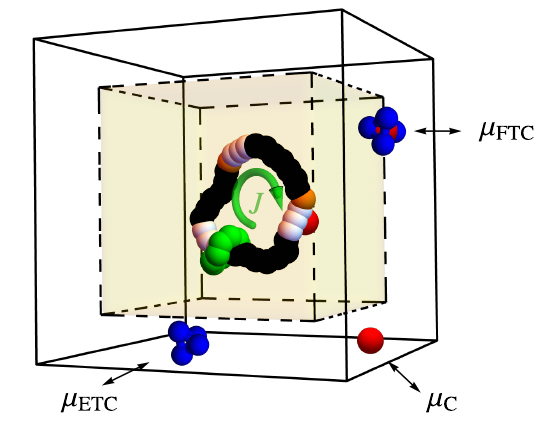}
    \hspace{-0.05in}
    \includegraphics[trim=0.39cm 0 0 0,height=0.26\textwidth]{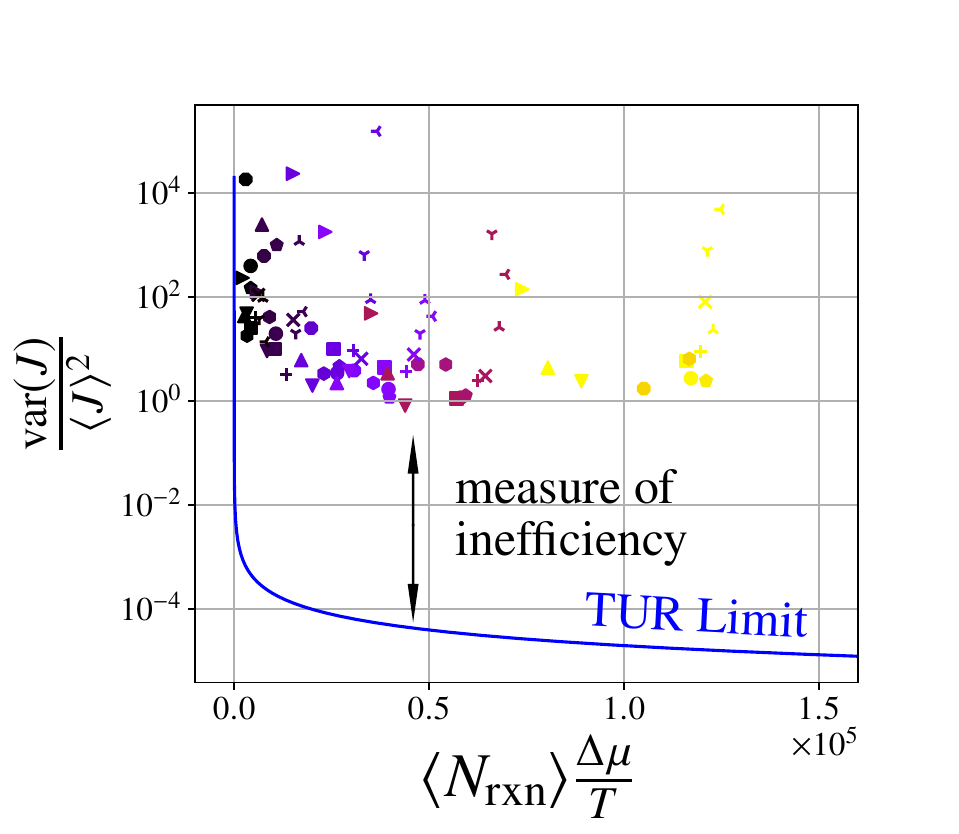}
    \hspace{-0.1in}
    \includegraphics[trim=0.35cm 0 0 0,height=0.26\textwidth]{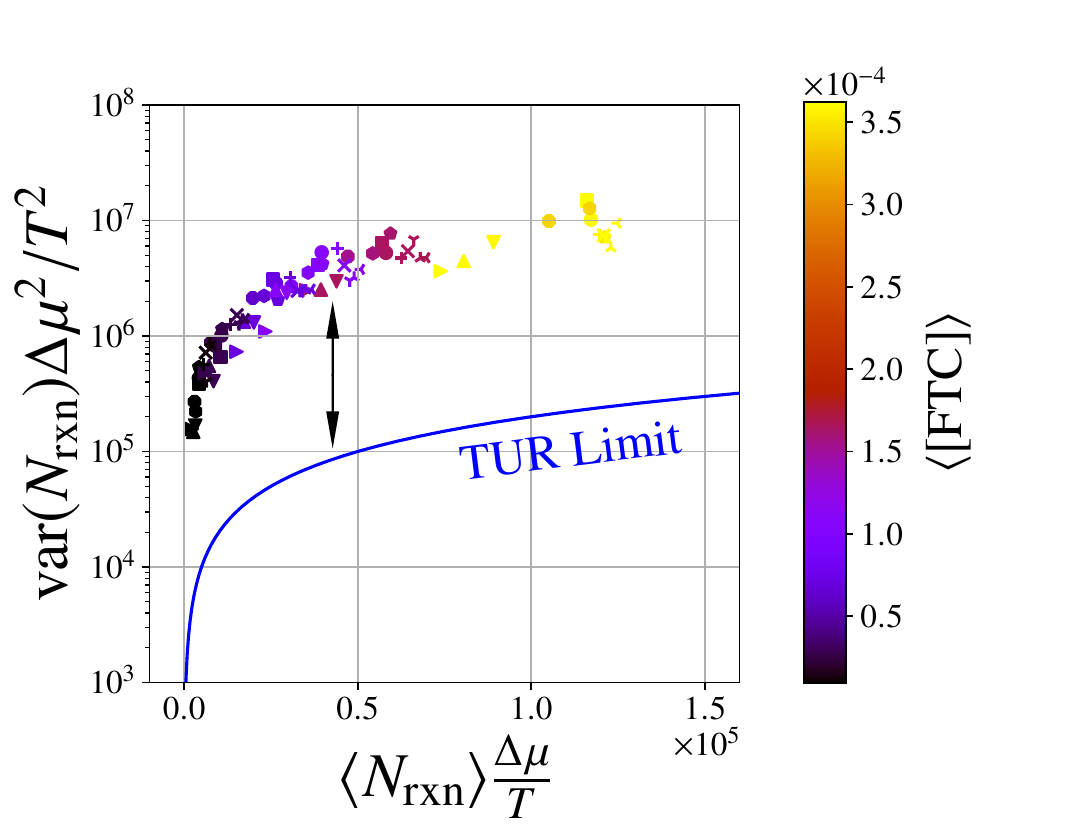}
    };
    \begin{scope}[x={(image.south east)},y={(image.north west)}]
        \node[] at (-0.02, 0.95) {(a)};
        \node[] at (0.32, 0.95) {(b)};
        \node[] at (0.635, 0.95) {(c)};
    \end{scope}
\end{tikzpicture}

\caption{(\emph{a}) The simulated motor model consisting of a green shuttling ring that diffuses around a larger ring and preferentially binds at orange sites. The net decomposition of \(N_{\rm rxn}\) full tetrahedral clusters (FTC, red and blue) into empty tetrahedral clusters (ETC, blue) and free particles (C, red) couples to the motion of the shuttling ring and induces a time-integrated current \(J\), which measures the net number of cycles in the clockwise direction. The coupling between the current and fuel decomposition results from attractive energetic interactions between the FTC clusters and the white binding sites, which catalyze fuel decomposition. 
By tuning the chemical potentials of the three species, \(\mu_\mathrm{FTC}\), \(\mu_\mathrm{ETC}\), and \(\mu_\mathrm{C}\), we impose a nonequilibrium driving force \(\Delta \mu = \mu_\mathrm{FTC} - \mu_\mathrm{ETC} - \mu_\mathrm{C}\) via a grand canonical Monte Carlo procedure that tends to inject FTC into and remove ETC and C from the system at temperature \(T\)~\cite{albaugh2022simulating}.
Plotted in (\emph{b}) and (\emph{c}) are comparisons of the data with the TUR limits Eqs.~\eqref{eq:tur} and~\eqref{eq:tur_ent} (\emph{c}), respectively, for a family of catenane motors with varying numbers of binding and catalytic sites, friction, and backbone rigidity, denoted with different symbols (see Supporting Information (SI) Sec.\ I for details).
We choose to express the \(x\) axis of each plot in terms of the typical dissipation \(\langle \Sigma \rangle = \langle \Sigma_{\rm rxn} \rangle = \langle N_{\rm rxn}\rangle \Delta \mu / T\).
  The color bar denotes the concentration of FTC molecules, which is controlled by the associated chemical potential.
}
  \label{fig:sim_tur}
\end{figure*}

How efficiently a motor approaches the TUR limit is a complicated function of a high-dimensional design space.
Depending on the interaction strengths between components of the motor, the motor can range from a precise machine to a dud, scenarios that we illustrate with molecular simulations of a family of catenane motors.
Our main results offer quantitative measures of the degree of saturation of the TUR bound in these catenane motors.
While the large design space yields motors with precisions varying over several orders of magnitude, we can anticipate how close each motor will get to the TUR bound by knowing just four physical properties: the chemical potential driving the motor, the rate of fuel decomposition, the coupling between fuel decomposition and motor motion, and the rate of undriven motor motion.
We derive such a simplification using a minimal Markov model, and illustrate that the resulting expressions, Eqs.~\eqref{eq:entnoweq} and \eqref{eq:curnoweq}, are instructive in explaining the performance of the more complicated molecular dynamics simulations of catenane motors.
We further translate Eq.~\eqref{eq:curnoweq} into a biophysical context to explain why biological motors can operate orders of magnitude closer to the TUR bound than presently demonstrated artificial ones.
Understanding how closely artificial molecular motors can saturate these fundamental TUR bounds highlights inefficiencies in their design and provides insight into how we might improve future artificial motors.

\section*{Molecular Motor Simulations}
We use a classical particle-based model of a molecular motor~\cite{albaugh2022simulating} inspired by the first synthetic, autonomous, chemically fueled molecular motor~\cite{wilson2016autonomous}.
As shown in Fig.~\ref{fig:sim_tur}a, the catenane motor consists of two interlocked rings.
The smaller shuttling ring traverses the larger ring by diffusion.
Both rings are composed of particles that are held together by nearest-neighbor bonds.
The larger ring contains a number of motifs, each consisting of a binding site (shown in orange) directly adjacent to a catalytic site (shown in white).
Binding sites act as potential-energy wells for the shuttling ring, whereas catalytic sites facilitate the decomposition of fuel molecules present in the bulk.

The fuel is represented by a full tetrahedral cluster (FTC) that consists of two components: an empty tetrahedral cluster (ETC) and a central particle (C), which is kinetically trapped within the ETC.
Whereas the motor is confined to an inner volume of the simulation box, the FTC, ETC, and C are free to pass between both inner and outer volumes.
The fuel decomposes when the C escapes from the FTC to form the ETC and C.
After a fuel decomposition event, the C may remain on the catalytic site for some time, resulting in steric hindrance that blocks the motion of the shuttling ring.
Further, the presence of the shuttling ring at a binding site inhibits catalysis at the proximal catalytic site, again by steric hindrance.
The resulting kinetic asymmetry couples the ring-and-fuel system and creates an information ratchet that gates the natural diffusion of the shuttling ring in a preferred direction~\cite{albaugh2022simulating, astumian2016running,amano2022insights, albaugh2023sterically}.
Modifying the spacing between motifs on the large ring can even allow for control over the shuttling ring's preferred direction~\cite{albaugh2023sterically}.

The whole system undergoes Langevin dynamics interspersed with periodic grand canonical Monte Carlo (GCMC) moves in the outer volume that set up a nonequilibrium chemical potential gradient.
The chemical potentials of FTC, ETC, and C in the outer volume, \(\mu_{\mathrm{FTC}}\), \(\mu_{\mathrm{ETC}}\), and \(\mu_{\mathrm{C}}\) respectively, are fixed via GCMC moves \cite{albaugh2022simulating, frenkel2001understanding,chempath2003two}, which act as chemostats. 
By setting \(\mu_{\mathrm{FTC}}\) high and \(\mu_{\mathrm{ETC}}\) and \(\mu_{\mathrm{C}}\) low, we induce favorable conditions for fuel decomposition and generate a chemical potential gradient
\begin{equation}
    \Delta\mu = \mu_\mathrm{FTC}-\mu_\mathrm{ETC} - \mu_\mathrm{C}
\end{equation}
that tends to introduce the FTC into the system and simultaneously remove the ETC and C from it.
This simulation setup allows us to perform numerical experiments with varying motor configurations, pair potentials, chemical potentials, frictions, and temperatures.
Results from the array of simulations are all plotted together in Fig.~\ref{fig:sim_tur} with the color of the plot markers reflecting the fuel concentration and the symbols representing motors with different characteristics, as described further in the SI.

\section*{Dissipation}
Each step in the simulation is microscopically reversible so that the dissipation can be rigorously computed as the entropy production in the reservoirs.
Due to local detailed balance with those reservoirs, the entropy produced during a single trajectory \(\mathbf{x}\) studied for an observation time \(t_\mathrm{obs}\) can equivalently be measured by the statistical irreversibility of the trajectory.
From that perspective, the entropy production associated with the trajectory, \(\Sigma(\mathbf{x})\), is expressed as the log-ratio of the probability of observing \(\mathbf{x}\) to its time-reversed analogue \(\tilde{\mathbf{x}}\) \cite{seifert2005entropy}, 
\begin{equation}
    \Sigma(\mathbf{x}) = k_{\mathrm{B}} \ln\frac{P(\mathbf{x})}{P(\tilde{\mathbf{x}})}.
    \label{eq:entrdefn}
\end{equation}
An individual trajectory is stochastic because of random thermal noise from the Langevin dynamics and the addition and removal of the FTC, ETC, and C via GCMC moves.
These factors combine multiplicatively to form \(P(\mathbf{x})\) and \(P(\tilde{\mathbf{x}})\), allowing \(\Sigma\) to be computed directly from simulations via Eq.~\eqref{eq:entrdefn}.
That total entropy production can be decomposed into components arising from the Langevin dynamics~\cite{sivak2013using,sivak2014time} and the chemostats.
The decomposition, shown explicitly in the SI, illustrates that the average entropy production can equivalently be computed from the entropy production of the reaction alone: \(\langle \Sigma\rangle = \langle \Sigma_{\rm rxn} \rangle = \langle N_{\rm rxn} \rangle\Delta \mu/ T\).

\section*{The Current and Its Precision}
To analyze the TURs, we must compare this typical entropy production with the fluctuations of time-integrated currents.
Two such currents are \(N_{\rm rxn}\) and the physical current that counts the net displacement of the shuttling ring around the large ring in its preferred direction, a current we call \(J\).
Means and variances of both currents are readily extracted by sampling simulated trajectories.
In the case of \(N_{\rm rxn}\) this merely requires that one count how many net FTC decompositions have occurred in a given time \(t_{\mathrm{obs}}\).
For \(J\), one counts the net number of large-ring particles that the shuttling ring passes \(\Delta n\) in that same \(t_{\mathrm{obs}}\).

Figs.~\ref{fig:sim_tur}b and~\ref{fig:sim_tur}c show the precision of the currents for a variety of model configurations and operating conditions, and how they compare to the TURs for \(J\) and \(N_{\rm rxn}\), Eqs.~\eqref{eq:tur} and~\eqref{eq:tur_ent}, respectively.
Both currents and their precisions depend strongly on \(\mu_\mathrm{FTC}\).
In alignment with the governing TURs, increasing the FTC concentration generally drives more current and decreases its relative variance.
However, for both \(J\) and \(N_{\rm rxn}\), the magnitude of the precision is far from the TUR bound---five orders of magnitude for \(J\), and two for \(N_{\rm rxn}\)---implying stark inefficiency that we subsequently rationalize.
Interestingly, the precision of \(N_{\rm rxn}\) appears to collapse onto a single curve in Fig.~\ref{fig:sim_tur}c, suggesting that the data are governed by some TUR-like relationship between the mean and variance.

\begin{figure}[t!]
    \centering
    \includegraphics[width=0.45\textwidth]{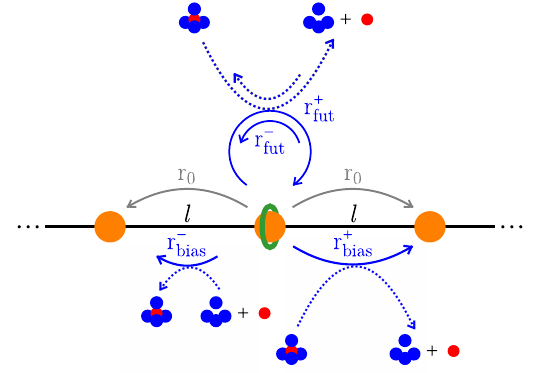}
    \caption{
    Schematic of a minimal model of a loosely coupled motor.
    Each state in the model represents a binding site on the large ring separated by a distance \(l\).
    Movement between these states corresponds to the shuttling ring hopping between binding sites.
    There are three different types of dynamic events that can occur: biased motion with rates \(r_\mathrm{bias}^\pm\), unbiased motion with rate \(r_0\), and futile fuel decomposition with rates \(r_\mathrm{fut}^\pm\).
    As illustrated, certain events are coupled to the decomposition or reconstitution of fuel, and others are not.
    Explicit expressions for the rates of each event are provided in the main text.}
    \label{fig:minmodelsketch}
\end{figure}

\section*{A Minimal Model of a Loosely Coupled Motor}
To pursue the idea that appropriately rescaled data would collapse onto a TUR-like relationship, we introduce a minimal Markov model for which we could analytically compute such a rescaling.
This minimal model should share the essential structure of the simulated catenane motors.
For example, the simulations reveal a loosely coupled motor; only some of the fuel decompositions actually result in the directed motion of the shuttling ring.
The loose coupling motivates us to introduce a minimal Markov model that involves three distinct events: fuel-coupled ring motion, fuel-decoupled ring motion and futile fuel reactions.
In terms of this simple model, we are able to quantitatively reproduce the features discussed in Fig.~\ref{fig:sim_tur}, which allows us to identify the factors that contribute to the looseness of the bounds. 
The model also enables straightforward analyses of design strategies for altering the motor to tighten these bounds.

The Markov model, illustrated in Fig.~\ref{fig:minmodelsketch}, consists of a shuttling ring that hops along an infinite track with binding sites separated by a distance \(l\).
This infinite track can be thought of as the result of unfurling the large ring of the catenane into a linear track with periodic replicas.
We focus on the coupling between the motion of the shuttling ring and the fuel transformation events, whereby fuel transformation refers both to the decomposition of the FTC to form the waste products ETC and C and to the reconstitution of the FTC from the ETC and C.
At each moment in time, one of three dynamical events can take place: fuel transformation coupled to biased motion with rates \(r_\mathrm{bias}^\pm\), futile fuel transformation not coupled to motion with rates \(r_\mathrm{fut}^\pm\), and unbiased motion not coupled to fuel transformation with rate \(r_0\).
The superscript \(+\) denotes decomposition of fuel, whereas the superscript \(-\) denotes reconstitution of fuel.
Fuel transformation is coupled to biased motion in the sense that the decomposition of one unit of fuel causes the ring to move one hop in the favored direction, cf.~Fig.~\ref{fig:minmodelsketch}.
Therefore, the ring dynamics are modeled as a superposition of a symmetric random walker perfectly decoupled from fuel dynamics and an asymmetric random walk perfectly coupled to fuel dynamics. 
The degree of coupling is quantified as the proportion of the fuel transformation coupled to biased motion.

The rates of the Markov model can be recast in terms of parameters with clear physical meaning: \(\zeta\), the combined rate of unbiased movement events; \(\lambda\), the combined rate of fuel transformation events; \(\eta\), the proportion of fuel transformation coupled to biased motion; and \(\Delta \mu\), the change in free energy associated with the decomposition of a single FTC species in solution, as calculated for the particle simulations in the SI~\cite{shirts2008statistically}. 
For thermodynamic consistency, each forward move must have a reverse counterpart.
The ratio of rates for these pairs of moves is proportional to the exponential of the free energetic difference that drives the bias.
Because the ring has degenerate binding sites, microscopic reversibility requires \(r^+ = r^-\exp(\beta\Delta\mu)\).
Furthermore, by definition, the rate of biased motion must be \(\eta\lambda\), and that for futile fuel transformation must be \((1-\eta)\lambda\).
These constraints uniquely specify the rates \(r_\mathrm{bias}^+ = r_\mathrm{bias}^-e^{\beta\Delta\mu} = \eta\lambda/(1+e^{-\beta\Delta\mu})\), \(r_\mathrm{fut}^+ = r_\mathrm{fut}^-e^{\beta\Delta\mu} = (1-\eta)\lambda/(1+e^{-\beta\Delta\mu})\), and \(r_0 = \zeta/2\).
The particle simulation operates under conditions in which \(\beta\Delta\mu \gg 1\) and hence \(r^+ \gg r^-\).

\section*{Current Fluctuations in the Minimal Model}
We let \(N_0^+\), \(N_0^-\), \(N_\mathrm{fut}^+\), \(N_\mathrm{fut}^-\), \(N_\mathrm{bias}^+\), and \(N_\mathrm{bias}^-\) be the number of forward (+) and reverse (\(-\)) moves accrued in time \(t_{\rm obs}\) due, respectively, to unbiased motion, futile fuel decomposition, and fuel decomposition coupled to biased motion.
With the Markovian assumption, these variables are all Poisson distributed with parameters equal to the corresponding rates multiplied by \(t_\mathrm{obs}\).
Both the shuttling-ring and fuel-decomposition currents can be expressed in terms of those Poisson variables: \(J = (N^+_{\rm bias} - N^-_{\rm bias} + N_0^+ - N_0^-) l\) and \(N_\mathrm{rxn} = N^+_{\rm bias} - N^-_{\rm bias} + N_{\rm fut}^+ - N_{\rm fut}^-\), where \(l\) is the length scale separating the binding sites in Fig.~\ref{fig:minmodelsketch}.
Making use of the fact that a Poisson-distributed \(N\) with the parameter \(r\cdot t_{\rm obs}\) has \(\langle N\rangle = \mathrm{var}(N) = r\cdot t_\mathrm{obs}\), the mean and variance of \(J\) and \(N_{\rm rxn}\) in the Markov model can be calculated.
For \(N_{\rm rxn}\), that calculation yields
\begin{align}
  \langle N_\mathrm{rxn} \rangle &= \lambda t_{\rm obs} \tanh \Big(\frac{\beta \Delta \mu}{2}\Big),
 \label{eq:aveSigma}\\
 \mathrm{var}(N_\mathrm{rxn}) &= \lambda t_{\rm obs},
 \label{eq:varSigma}
\end{align}
which can be rearranged into
\begin{equation}
  \mathrm{var}(N_{\rm rxn}) f\left(\frac{\beta \Delta \mu}{2}\right) = \frac{2 }{\beta \Delta \mu} \langle N_{\rm rxn} \rangle,  
  \label{eq:entnoweq}
\end{equation}
an \emph{equality} resembling the TUR \emph{inequality}, Eq.~\eqref{eq:tur_ent}.
Here, the saturation of the TUR inequality is determined by the scaling factor \(f(x) = \tanh (x)/x\), which tends to one as $\beta \Delta \mu$ vanishes.
For the Markov model of Fig.~\ref{fig:minmodelsketch}, we see that the tightness of the \(N_{\rm rxn}\) fluctuations relative to its TUR bound is exclusively regulated by \(\beta \Delta \mu\), the dimensionless free energy of decomposition.

Repeating the Poisson analysis for \(J\) gives the mean and variance
\begin{align}
  \langle J \rangle &= \eta \lambda \tanh\Big(\frac{\beta \Delta \mu}{2} \Big)t_\mathrm{obs}l,
 \label{eq:aveJ}\\
  \mathrm{var}(J) &= (\eta \lambda + \zeta) t_\mathrm{obs}l^2.
 \label{eq:varJ}
\end{align}
Rearrangement yields
\begin{equation}
  \frac{\mathrm{var}(J)}{\langle {J} \rangle^2} f\left(\frac{\beta \Delta\mu}{2}\right) g(\eta,\zeta,\lambda) = \frac{2k_\mathrm{B}}{\langle \Sigma \rangle },
  \label{eq:curnoweq}
\end{equation}
an \emph{equality} resembling the current TUR \emph{inequality}, Eq.~\eqref{eq:tur}.
Now, the saturation of the TUR for \(J\) is determined by \(f(\beta \Delta\mu/2)\) and \(g(\eta,\zeta,\lambda) =  \eta^2 \lambda/(\eta \lambda + \zeta)\).
The same factor \(f(\beta \Delta\mu/2)\) that appeared in Eq.~\eqref{eq:entnoweq} reflects that the minimal model's saturation of the TUR for fuel decomposition is necessary but insufficient to also saturate the TUR for \(J\).

The TUR equalities, Eqs.~\eqref{eq:entnoweq} and~\eqref{eq:curnoweq}, only rigorously apply to the minimal Markov model.
It is not obvious that they would provide direct insight into the fluctuations of the more complicated simulations.
The minimal Markov model imagines independently varying \(\zeta\), \(\eta\), \(\lambda\), and \(\Delta \mu\), but modifying motor interactions in the molecular dynamics simulations simultaneously changes all four parameters.
Remarkably, we show that for a broad array of motor designs, both Eqs.~\eqref{eq:entnoweq} and~\eqref{eq:curnoweq} are able to describe the simulation fluctuations if the Markov model parameters \(\zeta\), \(\eta\), and \(\lambda\) are replaced by effective values \(\zeta_{\rm eff}\), \(\eta_{\rm eff}\), and \(\lambda_{\rm eff}\) extracted from the simulations, as described in the SI.

\begin{figure}
\centering
\begin{tikzpicture}
\node[anchor=south west,inner sep=0] (image) at (0,0) {
\includegraphics[width=0.44\textwidth]{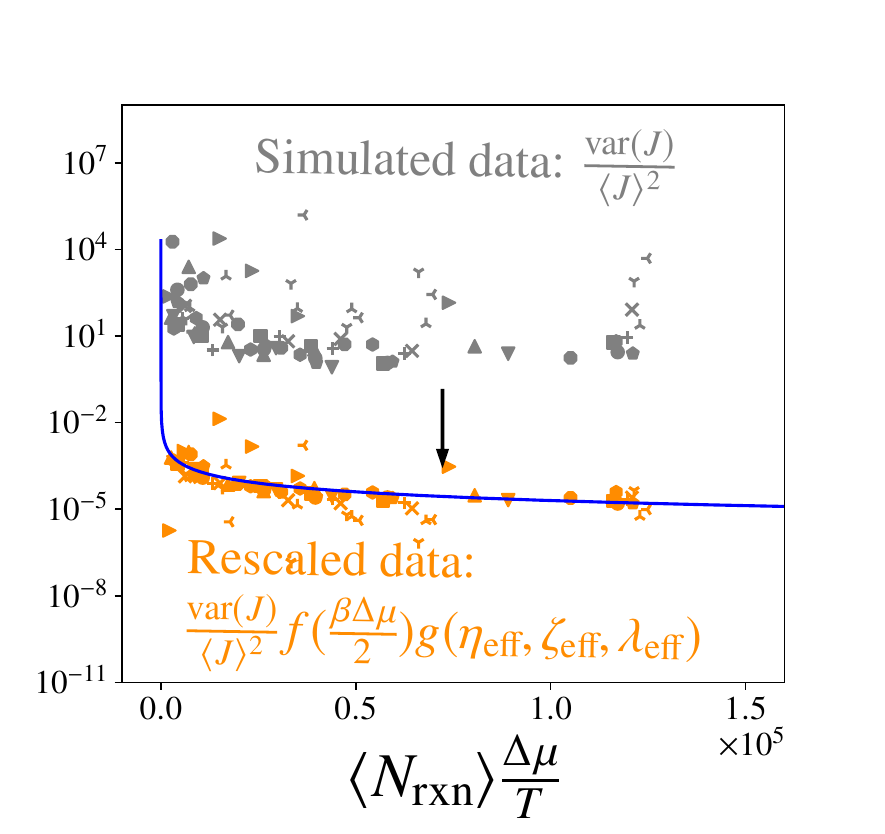}
};
    \begin{scope}[x={(image.south east)},y={(image.north west)}]
        \node[] at (-0.02, 0.95) {(a)};
    \end{scope}
\end{tikzpicture}
\begin{tikzpicture}
\node[anchor=south west,inner sep=0] (image) at (0,0) {
\includegraphics[width=0.44\textwidth]{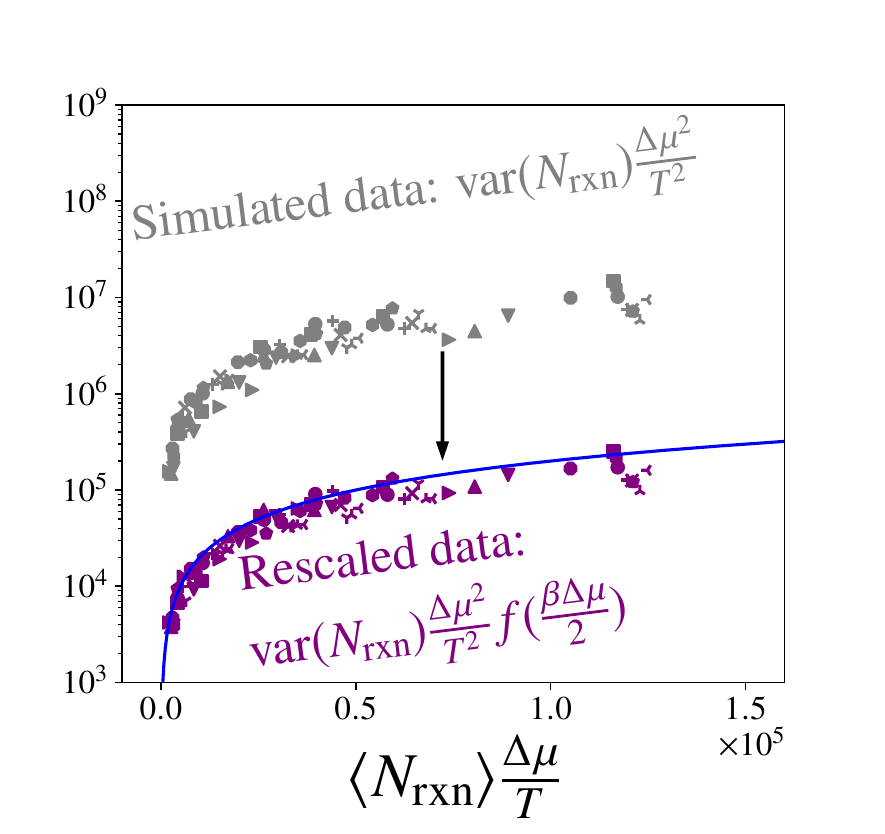}
};
    \begin{scope}[x={(image.south east)},y={(image.north west)}]
        \node[] at (-0.02, 0.95) {(b)};
    \end{scope}
\end{tikzpicture}
\caption{
  Plots showing the rescaling of Eqs.~\eqref{eq:Jrescaled} and~\eqref{eq:Nrescaled} applied to the simulated \(J\) and \(N_{\rm rxn}\) current fluctuations from molecular dynamics simulations of catenane motors.
  That the rescaled fluctuations lie on the curves of the TUR bounds indicates that the analytical scaling factors quantify how far the fluctuations are from the TUR bounds.
}
\label{fig:compare}
\end{figure}

To the extent that the minimal Markov model captures the simulated fluctuations, Eqs.~\eqref{eq:entnoweq} and~\eqref{eq:curnoweq} therefore imply that the TUR inequalities would be converted into an equalities if we rescaled variances:
\begin{align}
  \text{var}(J) &\rightarrow \text{var}(J) f\left(\frac{\beta \Delta \mu}{2}\right)g(\eta_{\rm eff}, \zeta_{\rm eff}, \lambda_{\rm eff})
  \label{eq:Jrescaled}\\
  \text{var}(N_{\rm rxn}) &\rightarrow \text{var}(N_{\rm rxn}) f\left(\frac{\beta \Delta \mu}{2}\right).
  \label{eq:Nrescaled}
\end{align}
Fig.~\ref{fig:compare} shows the result of applying that rescaling on the simulation data.
The rescaled fluctuations are effectively mapped onto the TUR curves, implying that the rescaling factors \(f\) and \(g\) are measures of how close the fluctuations get to saturating the TUR inequalities.
Crucially, because these \(f\) and \(g\) factors are expressed in terms of physically interpretable parameters, it allows us to address how those physical quantities (the chemical potential driving the motor, the rate of fuel decomposition, the coupling between fuel decomposition and motor motion, and the rate of undriven motor motion)  impact the degree of TUR saturation.

\section*{Discussion}
The TUR has attracted great interest because the inequality connecting motor precision with thermodynamics is generic, but the generality of the result can obscure the fact that far-from-equilibrium machines can operate far from the TUR bound.
Indeed, in our simulations of an artificial catenane motor, we have shown that the fluctuations in current deviate from the TUR bound by 5---6 orders of magnitude.
Even when the TUR is loose, we have here demonstrated that we can closely approximate the fluctuations in terms of the TUR via the rescaling of Eq.~\eqref{eq:Jrescaled} and~\eqref{eq:Nrescaled}.
This connection allows us to attribute the observed deviations from the TUR to both the large free energy difference driving fuel decomposition (\(\beta\Delta\mu\) ranged from \(74\) to \(118\) for our numerical experiments) and the low current from lack of coupling between fuel decomposition and ring movement (\(\eta_{\rm eff}\) ranged from 0 to 0.11).
Both \(\Delta \mu\) and coupling are clearly important factors in considering motor performance \cite{borsley2022tuning, borsley2022chemical}, and \(f(\beta \Delta \mu / 2\)) and \(g(\eta_{\rm eff}, \eta_{\rm eff}, \lambda_{\rm eff})\) now quantitatively highlight their impact on current fluctuations in these catenane motors.

From the rescaling, we see that minimizing the deviation from the TUR requires a combination of tight coupling and low entropy production from fuel decomposition.
TUR saturation simultaneously demands perfect coupling (\(\eta_{\rm eff} = 1\)), a vanishing driving force (\(\Delta \mu \to 0\)), and no unbiased movement (\(\zeta = 0\)).
It is suggestive to compare with biophysical motors driven by ATP hydrolysis, which for physiological conditions means \(\Delta \mu \approx 20 k_{\rm B} T\) and \(f(\beta \Delta \mu / 2) \approx 0.1\).
That \(\Delta \mu\) places a limit on the maximal achievable precision, but unlike our catenane simulations, the \(g\) term will have a limited role in most biophysical situations.
The difference is that the biophysical motors typically benefit from tight mechanical coupling~\cite{soga2017perfect} between fuel decomposition and mechanical motion, so that \(\eta \approx 1\), \(\zeta \ll 1\), and \(g(\eta_{\rm eff}, \zeta_{\rm eff}, \lambda_{\rm eff}) \approx 1\).
Without realizing similar tight coupling in synthetic motors, it will be hard to engineer them to reach the precision of their biophysical counterparts.

\section*{Acknowledgments}
We gratefully acknowledge insightful discussions with Emanuele Penocchio.
Research reported in this publication was supported by the Gordon and Betty Moore Foundation through Grant No.\ GBMF10790.

\bibliography{biblio.bib}

\renewcommand{\thefigure}{S\arabic{figure}}
\setcounter{figure}{0}
\renewcommand{\thetable}{S\arabic{table}}
\setcounter{table}{0}
\renewcommand{\theequation}{S\arabic{equation}}
\setcounter{equation}{0}

\onecolumngrid
\section*{Supporting Information}

\section{Simulation Details}
\label{sec:deets}

\begin{figure}[h]
\centering
\includegraphics[width=0.45\textwidth]{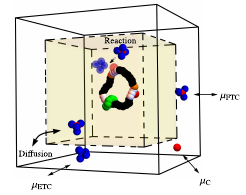}
\caption{
The motor consists of a smaller shuttling ring (green) that diffuses around a large ring.
The shuttling ring preferentially resides at binding sites (orange) located next to catalytic sites (white), which catalyze a decomposition reaction.
The decomposition reaction of an FTC (blue and red) into an ETC (blue) and C (red), can leave a C bound to the large ring, impeding the diffusion of the shuttling ring.
The C can also naturally unbind and diffuse away.
By coupling these processes with GCMC moves that maintain a high concentration of FTC and a low concentration of ETC and C, the shuttling ring's diffusion can be biased in a preferred direction as a result of the gradient in chemical potential caused by the nonequilibrium concentrations.
The motor is constrained to the inner box (yellow) by a wall potential.
GCMC only occurs in the outer volume (white).
FTC, ETC, and C are free to pass between the inner and outer volumes, resulting in nonequilibrium steady-state concentrations and a constant net movement of the shuttling ring.}
\label{fig:system}
\end{figure}

We summarize the motor model below; a more detailed description of the model and methods are found in~\cite{albaugh2022simulating}.
Our motor model consists of two interlocked rings based on a synthetic, autonomous, chemically fueled molecular motor~\cite{wilson2016autonomous}.
As shown in Fig.~\ref{fig:system}, the smaller shuttling ring (green) diffuses around the larger ring.
The larger ring consists of binding sites (orange), which attract the shuttling ring, and white particles that catalyze a classical ``reaction''.
The black particles are purely volume-excluding and serve to complete the ring structure.
The classical ``reaction'' that fuels the motor is the decomposition of a full tetrahedral cluster (FTC) into an empty tetrahedral cluster (ETC) and free central particle (C).
The FTC consists of a tetrahedron (blue) bound along its edges with a free central particle (red) kinetically trapped in the center of the tetrahedron.
This trapped C particle is in an entropically and energetically unfavorable position.
To escape, it would need to deform the tetrahedron, leading to a metastable, high free energy state.
We choose our reaction coordinate as the distance \(d\) between the C particle and the center of mass of a tetrahedron (ETC).
If \(d < 0.25\) then the pair is in the FTC state.
If \(d > 0.8\) then the pair is in the ETC and C state.
For \( 0.25 \le d \le 0.8\) the pair is in a transition state.
Section~\ref{sec:free_en} discusses the energetics of this classical reaction further.

Attractive interactions between the catalytic sites and FTC facilitate decomposition of the FTC into ETC and C near the motor.
The catalyzed reaction leaves behind a C particle bound to the white catalytic site, which prevents diffusion of the shuttling ring.
If the shuttling ring is at a binding site, it will sterically hinder FTC from binding to the adjacent catalytic site.
However, for catalytic sites further from the shuttling ring, FTC is otherwise unhindered and can freely approach, decompose, and leave a blocking group on those sites.
As a result, blocking groups are more frequently created due to catalysis at sites far from the shuttling ring's present location, resulting in kinetic asymmetry and creating an information ratchet that preferentially gates the diffusion of the shuttling ring in the clockwise direction for large spacing between sites and in the counterclockwise direction for small spacing.
This directionality is driven by steric effects and is discussed in~\cite{albaugh2023sterically}.

A simple set of potentials govern particle interactions in the system.
Adjacent ring particles are bound by a finitely extensible nonlinear elastic (FENE) spring with potential
\begin{equation}
U_{\mathrm{FENE}} = \frac{1}{2} \kappa_{ij} r^{2}_{0,ij} \ln{ \left[1 - \left( \frac{r_{ij}}{r_{0,ij}} \right)^{2} \right]}.
\label{eq:fene}
\end{equation}
Here, \(\kappa_{ij}\) is the FENE force constant between particles \(i\) and \(j\), \(r_{0,ij}\) is the maximum extension, and \(r_{ij}\) is the current distance between the particles.
The tetrahedron particles are bound with harmonic springs, with potential
\begin{equation}
U_{\mathrm{harmonic}} = \frac{1}{2} k_{ij} r_{ij}^{2}.
\label{eq:harmonic}
\end{equation}
Here, \(k_{ij}\) is the harmonic force constant between the two particles \(i\) and \(j\).
Groups of three adjacent ring particles also have angular potentials to help maintain their circular shape, and these potentials take the form
\begin{equation}
U_{\mathrm{angle}} = \frac{1}{2} K_{ijk} \left(\theta_{ijk} - \theta_{0,ijk} \right)^{2},
\label{eq:angle}
\end{equation}
where \(\theta_{ijk}\) is the angle formed by the three particles \(i\), \(j\), and \(k\) in that order, \(\theta_{0,ijk}\) is the equilibrium angle, and \(K_{ijk}\) is the angular force constant.
These spring and angular interactions maintain the bonded structure of the rings and tetrahedrons. 
All particle pairs also interact through a 12-6 potential,
\begin{equation}
U_{\mathrm{LJ}} = 4\epsilon_{\mathrm{R},ij} \left(\frac{\sigma_{ij}}{r_{ij}} \right)^{12} - 4\epsilon_{\mathrm{A},ij} \left(\frac{\sigma_{ij}}{r_{ij}} \right)^{6},
\label{eq:LJ}
\end{equation}
with \(\epsilon_{\mathrm{R},ij}\) controlling the repulsion between particles \(i\) and \(j\), and \(\epsilon_{\mathrm{A},ij}\) controlling the attraction.
The effective pair radius is \(\sigma_{ij}\).
Each pair of particles is at least volume-excluding, with \(\epsilon_{\mathrm{R},ij} > 0\), but not all interactions have an attractive component given by \(\epsilon_{\mathrm{A},ij} \ge 0\).
The only attractive interactions are between the shuttling ring (green) and binding site (orange) particles and also between the FTC components (blue and red) and catalytic site (white) particles.
All of these interactions are time-independent; the motor is not driven by temporally modulating any potentials.

To place the motor into an out-of-equilibrium state and measure its current, we use the simulation setup depicted in Fig.~\ref{fig:system}.
The motor is confined to the inner simulation cube with the simple Lennard-Jones wall potential
\begin{equation}
U_{\mathrm{wall}}= 4 \epsilon_{\mathrm{wall}} \sum_{\alpha=x,y,z} \Bigg[ \left( \frac{\sigma_{\mathrm{wall}}}{r_{\alpha,i} - \frac{1}{2} L_{\mathrm{inner}}} \right)^{12} +\left( \frac{\sigma_{\mathrm{wall}}}{r_{\alpha,i} + \frac{1}{2} L_{\mathrm{inner}}} \right)^{12} \Bigg].
\label{eq:LJwall}
\end{equation}
Here, \( \epsilon_{\mathrm{wall}}\) and \(\sigma_{\mathrm{wall}}\) are wall potential energy and distance parameters, respectively, \(r_{\alpha,i}\) is the \(\alpha\)-component (\(\alpha = x\), \(y\), or \(z\)) of the position vector of particle \(i\), and \(L_{\mathrm{inner}}\) is the side length of the inner simulation cube.
The outer simulation cube has length \(L_{\mathrm{outer}}\) and is concentric with the inner cube and the origin.
We perform grand canonical Monte Carlo (GCMC) moves for FTC, ETC, and C in the outer box.
FTC, ETC, and C are unaffected by the wall potential and can pass freely between the inner and outer box (and also across the periodic boundaries of the outer box).
In this way, the GCMC moves can maintain homogeneous out-of-equilibrium concentrations throughout the simulation volume without having a direct effect on the motor dynamics as long as the time scale of FTC, ETC, and C diffusion is fast relative to reaction.
We define the external chemical potential \(\mu^{\prime}=\mu - A_{0}\) as the absolute chemical potential \(\mu\) less the Helmholtz free energy of a single copy of the species in isolation in the simulation volume (\(V = L_{\mathrm{outer}}^{3}\)).
By setting the external chemical potential of FTC \(\mu_{\mathrm{FTC}}^{\prime}\) high and the external chemical potentials of ETC \(\mu_{\mathrm{ETC}}^{\prime}\) and C \(\mu_{\mathrm{C}}^{\prime}\) low, our GCMC moves will tend to add FTC to and remove ETC and C from the outer box.
This allows us to hold the concentrations of these three species out of equilibrium and create a driving force for the motor with a high concentration of fuel (FTC) and waste (ETC and C).

Prior to starting motor simulations, we sampled configurations of single FTC and ETC clusters using Markov-chain Monte Carlo (MCMC).
We generated libraries of \(10^5\) structures for each species at the simulation temperature \(T\).
These libraries were used during motor simulations to generate initial configurations for inserted FTC and ETC clusters.
This was not necessary for C since C is a single isotropic particle without any internal potential energy or different configurations.
A GCMC insertion of an FTC and ETC begins by uniformly randomly selecting a structure from the respective canonical library.
The structure is then randomly rotated.
At this point, the structure or the C particle is then uniformly randomly placed in the outer box.
This procedure approximates selecting a species from a Boltzmann ensemble with the inclusion of an indistinguishability factor.
The generation probability for producing this configuration of \(N+1\) molecules of the species from the original configuration containing \(N\) molecules of the species is
\begin{equation}
P_{\mathrm{gen}}(N \to N + 1) = \frac{1}{N+1} \frac{e^{-\beta H_{0}}}{Z_{0}}.
\end{equation}
Here, \(H_{0} = U_{0} + K_{0}\) is the total internal energy of the inserted species consisting of the intramolecular potential energy \(U_{0}\) of the species and its kinetic energy \(K_{0}\).
We further have \(\beta = (k_{\mathrm{B}}T)^{-1}\).
The probability of accepting this insertion is
\begin{equation}
    P_{\mathrm{acc}}(N \to N + 1) = \min{\left[1, \frac{e^{-\beta(\Delta U - \mu^{\prime})}}{N+1} \right]}, 
\end{equation}
where \(\Delta U := U(\mathbf{r}^{N+1})-(U(\mathbf{r}^{N})+U_0)\) is the energy of solvation for the inserted species: the difference between the potential energy when the system and inserted molecule interact, \(U(\mathbf{r}^{N+1})\), and that when they do not interact, \(U(\mathbf{r}^{N})+U_0\).
After calculating the acceptance probability, the move is accepted or rejected according to the Metropolis criterion.
If the move is accepted, the simulation carries on with the inserted particle; if it is rejected, the simulation reverts to the pre-insertion configuration.
To remove a copy of an FTC, ETC, or C from the simulation, we uniformly randomly select one and remove it if it is in the outer simulation box.
If the species is not within the outer box, we reject the move.
The generation probability of this move from a configuration of \(N+1\) of the species to \(N\) is then simply
\begin{equation}
P_{\mathrm{gen}}(N+1 \to N) = \frac{1}{N+1}.
\end{equation}
As before, we calculate the acceptance probability of this removal as
\begin{equation}
    P_{\mathrm{acc}}(N + 1\to N) = \min{\left[1, (N+1) e^{\beta \left( \Delta U - \mu^{\prime} \right)} \right]}.
\end{equation}
The move is subsequently accepted or rejected according to the Metropolis criterion.
If the move is accepted, the simulation carries on without the deleted particle; if it is rejected, the simulation reverts to the pre-deletion configuration.
In total, there are six possible GCMC moves: an insertion or deletion of each of the species FTC, ETC, and C.
Every 12,500 simulation time steps, we uniformly randomly select one of these six moves to attempt.

In between GCMC moves, the system dynamically evolves according to an underdamped Langevin equation for each particle \(i\),
\begin{equation}
\begin{aligned}
\dot{\mathbf{r}}_{i} & = \frac{\mathbf{p}_{i}}{m_{i}}
\\
\dot{\mathbf{p}}_{i} &= - \frac{\partial U(\{\mathbf{r})\}}{\partial \mathbf{r}_{i}} - \frac{\gamma}{m_{i}} \mathbf{p}_{i} + \boldsymbol{\xi}_{i},
\end{aligned}
\label{eq:langevin}
\end{equation}
where the potential energy \(U\) is a function of all positions \(\{\mathbf{r}\}\), \(\mathbf{r}_{i}\) is the position of particle \(i\), \(\mathbf{p}_{i}\) is the momentum of particle \(i\), \(m_{i}\) is the mass of particle \(i\), \(\gamma\) is the friction coefficient, and \(\boldsymbol{\xi}\) is a white noise satisfying \(\left<\boldsymbol{\xi}_{i}\right> = \mathbf{0}\) and \(\left<\boldsymbol{\xi}_{i}(t) \boldsymbol{\xi}_{j}(t')\right> = 2 \gamma k_{\rm B} T \delta(t - t') \delta_{ij} \mathbf{I}\), where \(\mathbf{I}\) is the identity matrix.
We numerically integrate the underdamped Langevin equation by breaking the simulation into time steps of size \(\Delta t\).
This discretization is stated explicitly in Eq.~\eqref{eq:forward} and described in more detail in Sec.~\ref{sec:derive}.
Unless otherwise stated, we used a time step of \(\Delta t = 0.002\) and ran trajectories for \(5 \times 10^{8}\) time steps with 50 independent trials to collect statistics.
To ensure steady states, the first \(1 \times 10^8\) steps were discarded and analyses were performing on the last \(4 \times 10^8\) steps.
We used a smaller time step than previous studies~\cite{albaugh2022simulating,albaugh2023sterically} to reduce the effect of integrator entropy production due to a finite time step~\cite{sivak2013using}.
A standard simulation uses \(\beta = 2\), \(\gamma = 0.5\), and a motor with two binding site/catalytic site motifs on opposite ends of the large ring.
Unless otherwise stated, simulations used \(\mu_{\mathrm{ETC}}^{\prime}=\mu_{\mathrm{C}}^{\prime}=-10\).
To study a range of FTC concentrations, we used \(\mu_{\mathrm{FTC}}^{\prime}=-1, -0.5, 0,0.25,0.5,1\).

To generate the data in Figs.\ 1, 3, \ref{fig:S3}, \ref{fig:sim_cat_tur}, and \ref{fig:times}, we performed base simulations with \(\mu_{\mathrm{ETC}}^{\prime} = \mu_{\mathrm{C}}^{\prime} = -10\), friction \(\gamma=0.5\), and two evenly spaced binding and catalytic motifs on the large ring. These dynamics are plotted with \(\Circle\).
From these base parameters, we varied the friction to \(\gamma=1\) (\( \pentagon \)), \(\gamma=5\) (\( \varhexagon \)), and \(\gamma=10\) (\( \octagon \)).
We varied the waste chemical potentials to \(\mu_{\mathrm{ETC}}^{\prime} = \mu_{\mathrm{C}}^{\prime} = -3\) (\( \medtriangleup \)), \(\mu_{\mathrm{ETC}}^{\prime} = \mu_{\mathrm{C}}^{\prime} = -1\) (\( \medtriangledown \)), and \(\mu_{\mathrm{ETC}}^{\prime} = \mu_{\mathrm{C}}^{\prime} = 0\) (\( \medtriangleright \)).
We varied the number of binding and catalytic motifs evenly spaced around a large ring of fixed size to (\( \medwhitestar \)), \(N_{\mathrm{motif}}=3\) (\(+\)), \(N_{\mathrm{motif}}=4\) (\( \times \)), \(N_{\mathrm{motif}}=5\) (\(  \Ydown  \)), \(N_{\mathrm{motif}}=6\) (\(  \Yup  \)), and \(N_{\mathrm{motif}}=7\) (\(  \Yleft  \)).
We also fixed the large ring in place (\( \medsquare \)).
Colors represent external chemical potential for FTC \(\mu_{\mathrm{FTC}}^{\prime}\), where higher \(\mu_{\mathrm{FTC}}^{\prime}\) corresponds to higher FTC concentration.

\section{Derivation of Entropy Production Terms}
\label{sec:derive}

We calculate the entropy production associated with our motor simulations from first principles at the level of stochastic trajectories.
For convenience, we work with the extensive entropy production \(\Sigma = \sigma t_\mathrm{obs}\).
The entropy produced along a trajectory of total time \(t_{\mathrm{obs}} = N_{\mathrm{steps}} \Delta t\) is proportional to the logarithm of the ratio of the probability of the forward trajectory \(\mathbf{x}\) to its time reversal \(\tilde{\mathbf{x}}\):
\begin{align}
    \frac{\Sigma_{t_\mathrm{obs}}}{k_\mathrm{B}} &= \ln{\left[ \frac{P(\mathbf{x})}{P(\tilde{\mathbf{x}})} \right]}\\
        &= \ln{\left[ \frac{P(\mathbf{x}(0))}{P(\mathbf{x}(t_{\mathrm{obs}}))} \right]} + \ln{\left[ \frac{P(\mathbf{x}|\mathbf{x}(0))}{P(\tilde{\mathbf{x}}|\mathbf{x}(t_{\mathrm{obs}}))}  \right]}. \label{eq:entprod}
\end{align}
Eq.~\eqref{eq:entprod} involves two terms, respectively a `boundary' term and a `body' term.
The boundary term deals with the change in entropy associated with the change in the system's state at the start and end of the trajectory.
This term is not extensive in time and will vanish in the steady state, so we exclude it from consideration.
The body term deals with the change in entropy due to heat dissipated into the surroundings and particle exchange with the chemostats over the course of the trajectory. It is extensive in time and scales linearly with the trajectory length.

The probability \( P(\mathbf{x}|\mathbf{x}(0)) \) is the probability of observing a trajectory \(\mathbf{x}\) given an initial configuration \(\mathbf{x}(0)\).
Similarly, the probability \( P(\tilde{\mathbf{x}}|\tilde{\mathbf{x}}(t_{\mathrm{tobs}})) \) is the probability of observing a time-reversed trajectory \(\tilde{\mathbf{x}}\) given an initial configuration \(\tilde{\mathbf{x}}(t_{\mathrm{tobs}})\).
Our simulations are composed of stochastic Langevin dynamics interspersed with GCMC moves, which operate independently of each other and hence lead to a factorizable trajectory probability \(P(\mathbf{x}|\mathbf{x}(0))\).
The entropy production, which is logarithmic in this probability, can therefore be decomposed into two additive terms,
\begin{equation}
    \Sigma = \Sigma_{\mathrm{Langevin}} + \Sigma_{\mathrm{MC}}.
\end{equation}

Our simulations generate Langevin dynamics using the numerical VRORV algorithm \cite{sivak2014time}, so named for the order of substeps within a complete time step. The velocities `V' are updated to half-step values, followed by the positions `R'. Subsequently, the random numbers `O' (for Ornstein–Uhlenbeck) are updated. Finally, the positions `R' and the velocities `V' are respectively updated by another half-step. This concludes a complete step.

The updates specified above are described by the following set of equations for a particle with mass \(m\), position \(\mathbf{r}\) and momentum \(\mathbf{p}\).
\begin{align}
    \mathbf{p}_{i+1/4} &= \mathbf{p}_i - \frac{\Delta t}{2} \partial_\mathbf{r}U_i \\
    \mathbf{r}_{i+1/2} &= \mathbf{r}_i + \frac{\Delta t}{2m} \mathbf{p}_{i+1/4}\\
    \mathbf{p}_{i+3/4} &= \alpha\mathbf{p}_{i+1/4} + \sqrt{\frac{m}{\beta}(1-\alpha^2)}\boldsymbol{\eta}_{i+3/4}\\
    \mathbf{r}_{i+1} &= \mathbf{r}_{i+1/2} + \frac{\Delta t}{2m}\mathbf{p}_{i+3/4} \\
    \mathbf{p}_{i+1} &= \mathbf{p}_{i+3/4} - \frac{\Delta t}{2} \partial_\mathbf{r}U_{i+1}, \label{eq:forward}
\end{align}
where \(\gamma\) is the friction coefficient, \(\Delta t\) is the timestep, \(\beta = (k_\mathrm{B}T)^{-1}\) is the inverse temperature, \(\alpha = \exp(- \gamma \Delta t/m)\), and \(U = U(\{\mathbf{r}\})\) is the potential function for the system specified with the positions of all particles \(\{\mathbf{r}\}\).
The subscript \(i\) indexes the time steps.
The standard Gaussian noise vector \(\boldsymbol{\eta}\) has components that are identically and independently distributed from a standard Gaussian distribution with zero mean and unit variance. 
We note that, although underdamped Langevin dynamics were used in the simulation, the separation of timescales between the frictional and inertial forces within the system places the system in the overdamped regime, justifying the use of the overdamped TUR bounds in the analyses of the main text.

This procedure takes the set of particle positions and momenta \(\{\mathbf{r}, \mathbf{p}\}\) at time \(t\) and advances them to time \(t+\Delta t\).
The time-reversed procedure flips the sign of all time-antisymmetric quantities (\(\mathbf{p} \to -\mathbf{p}\)) and swaps all indices as \(i+\delta \to i+1-\delta\):
\begin{align}
    \mathbf{p}_{i+3/4} &= \mathbf{p}_{i+1} + \frac{\Delta t}{2} \partial_\mathbf{r}U_{i+1} \\
    \mathbf{r}_{i+1/2} &= \mathbf{r}_{i+1} - \frac{\Delta t}{2m} \mathbf{p}_{i+3/4}\\
    \mathbf{p}_{i+1/4} &= \alpha \mathbf{p}_{i+3/4} - \sqrt{\frac{m}{\beta}(1-\alpha^2)}\tilde{\boldsymbol{\eta}}_{i+1/4} \label{eq:etatilde}\\
    \mathbf{r}_i &= \mathbf{r}_{i+1/2} - \frac{\Delta t}{2}\mathbf{p}_{i+1/4} \\
    \mathbf{p}_i &= \mathbf{p}_{i+1/4} + \frac{\Delta t}{2}\partial_\mathbf{r}U_i. \label{eq:reverse}
\end{align}

Note that the noise \(\tilde{\boldsymbol{\eta}}\) required for the reverse trajectory will generally be different from that required for the forward trajectory. To denote this difference, we have marked the time-reversed noise with a tilde.
Solving for \(\tilde{\boldsymbol{\eta}}_{i+1/4}\) in Eq.~\eqref{eq:etatilde} gives
\begin{equation}
\tilde{\boldsymbol{\eta}}_{i+1/4} = \sqrt{\frac{\beta}{m}}\frac{1}{\sqrt{1-\alpha^2}} \left(\alpha\mathbf{p}_{i+3/4} - \mathbf{p}_{i+1/4}\right).
\label{eq:backrand}
\end{equation}
Eq.~\eqref{eq:backrand} allows us to calculate the random vector \(\tilde{\boldsymbol{\eta}}\) that would produce time-reversed dynamics simultaneously as we perform forward integration during simulation.
The component of entropy production attributed to the Langevin dynamics can be expressed in terms of the random vectors \(\boldsymbol{\eta}\) and \(\tilde{\boldsymbol{\eta}}\) as
\begin{equation}
    \Sigma_{\mathrm{Langevin}} = k_{\mathrm{B}} \ln{ \left[ \prod_{i=1}^{N_{\mathrm{steps}}} \prod_{j=1}^{N_{\mathrm{particles}}} \frac{ P(\boldsymbol{\eta}^{j}_{i+3/4}) }{ P(\tilde{\boldsymbol{\eta}}^{j}_{i+1/4}) } \right] }. \label{eq:entrprodlangevin}
\end{equation}
Transforming the argument of the probability from the trajectory \(\mathbf{x}\) to the noises \(\{\boldsymbol{\eta}\}\) requires a Jacobian factor, which cancels out because we only consider ratios of probabilities.
\(N_{\mathrm{particles}}\) is the number of particles in the simulation box during the Langevin integration at time \(i \Delta t\).
This number can change during the simulation as GCMC moves add or remove particles.
The Langevin dynamics employs noises obeying a standard Gaussian distribution with zero mean and unit variance, so that the noise vector \(\boldsymbol{\eta}\) has distribution
\begin{equation}
    P(\boldsymbol{\eta}) = \frac{1}{(2\pi)^{n/2}}\exp(-\boldsymbol{\eta}^2/2),
\label{eq:noisedistr}
\end{equation}
where \(n\) is the dimensionality of \(\boldsymbol{\eta}\).
Substituting Eq.~\eqref{eq:noisedistr} into Eq.~\eqref{eq:entrprodlangevin}, we find that the entropy production is given by
\begin{equation}
   \Sigma_{\mathrm{Langevin}} = \frac{k_{\mathrm{B}}}{2}  \sum_{i=1}^{N_{\mathrm{steps}}} \sum_{j=1}^{N_{\mathrm{particles}}} \left( ({\tilde{\boldsymbol{\eta}}}^{j}_{i+1/4})^2 - ({\boldsymbol{\eta}}^{j}_{i+3/4})^2 \right).
    \label{eq:langevin_ent}
\end{equation}
We emphasize that the forward random numbers \(\boldsymbol{\eta}_{i+3/4}\) are drawn from a random number generator during the simulation, whereas the reverse random numbers \(\tilde{\boldsymbol{\eta}}_{i+1/4}\) are calculated from Eq.~\eqref{eq:backrand}.

The other component of the total entropy production \(\Sigma\) is that due to GCMC insertions and deletions, \(\Sigma_{\mathrm{MC}}\).
We construct this term from the probability of making insertion \((N \to N + 1)\) or deletion \((N+1 \to N)\) moves, where \(N\) is the number of molecules of the species being inserted or deleted: FTC, ETC, or C.
In terms of these probabilities, we have 
\begin{equation}
    \Sigma_{\mathrm{MC}} = k_{\mathrm{B}} \sum_{i=1}^{N_{\mathrm{GCMC}}} \Delta_{i} \ln{\left[ \frac{P(N_{i} \to N_{i} + 1)}{P(N_{i} + 1 \to N_{i})} \right]}.
\end{equation}
There are \(N_{\mathrm{GCMC}}\) total insertion and deletion moves, with \(\Delta=1\) for an insertion and \(\Delta=-1\) for a deletion.
An insertion move specifies the position \(\mathbf{r}\) and momentum \(\mathbf{p}\) of the new particle being added; a deletion move strips away that information.
An insertion or deletion move is performed if the move is proposed (with probability \(P_\mathrm{gen}\)) and subsequently accepted (with probability \(P_\mathrm{acc}\)), so that
\begin{equation}
\frac{P(N \to N + 1)}{P(N + 1 \to N)} = \frac{P_{\mathrm{gen}}(N \to N + 1)P_{\mathrm{acc}}(N \to N + 1) }{P_{\mathrm{gen}}(N + 1 \to N) P_{\mathrm{acc}}(N + 1 \to N)}.
\end{equation}
A proposed insertion move has Boltzmann probability weighted by an indistinguishability factor,
\begin{equation}
P_{\mathrm{gen}}(N \to N + 1) = \frac{1}{N+1} \frac{e^{-\beta H_{0}}}{Z_{0}},
\end{equation}
where \(H_{0} = U_{0} + K_{0}\) is the total internal energy of the inserted species consisting of the intramolecular potential energy \(U_{0}\) of the species and its kinetic energy \(K_{0}\).
The partition function \(Z_{0}\) is that of a lone molecule of this species in the volume \(V\) of the simulation box,
\begin{equation}
    Z_{0} = \frac{1}{h^{3n}} \int_{\mathbb{R}} \mathrm{d} \mathbf{p}^{n}  \int_{V} \mathrm{d} \mathbf{r}^{n}\,e^{-\beta H(\mathbf{r}^{n}, \mathbf{p}^{n})},
\end{equation}
where \(n\) is the number of particles in the species (5 for FTC, 4 for ETC, and 1 for C).
This partition function can be related to the standard Helmholtz free energy as:
\begin{equation}
A_{0}(V) = -k_{\mathrm{B}} T \ln{Z_{0}(V)}.
\label{eq:free_en}
\end{equation}
The proposed insertion move is accepted with the Metropolis probability
\begin{equation}
    P_{\mathrm{acc}}(N \to N + 1) = \min{\left[1, \frac{e^{-\beta(\Delta U - \mu^{\prime})}}{N+1} \right]}, \label{eq:metropolis}
\end{equation}
where \(\Delta U := U(\mathbf{r}^{N+1})-(U(\mathbf{r}^{N})+U_0)\) is the energy of solvation for the inserted species: the difference between the potential energy when the system and inserted molecule interact \((U(\mathbf{r}^{N+1}))\) and that when they do not interact \((U(\mathbf{r}^{N})+U_0)\). 
We denote by a prime \(\cdot^\prime\) the external, rather than the absolute, chemical potential of a species. The external chemical potential is the absolute chemical potential less the standard free energy,
\begin{equation}
    \mu^{\prime} = \mu - A_{0}.
\end{equation}
The value of \(\mu^\prime\) and \(\Delta U\) is species-dependent and will be determined by the species to be inserted or removed during a particular move.

A deletion is proposed with probability equivalent to that of selecting one of the present molecules of the chosen species for deletion:
\begin{equation}
P_{\mathrm{gen}}(N + 1 \to N) = \frac{1}{N+1}.
\end{equation}
The proposed deletion move is accepted again with the Metropolis probability
\begin{equation}
    P_{\mathrm{acc}}(N + 1\to N) = \min{\left[1, (N+1) e^{\beta \left( \Delta U - \mu^{\prime} \right)} \right]}.
\end{equation}

Combining these insertion and deletion probabilities with Eq.~\eqref{eq:free_en}, along with the identity \(\min\left[1,x\right]/\min\left[1,x^{-1} \right] = x\) for \(x > 0\), we obtain
\begin{equation}
    \frac{P(N \to N + 1)}{P(N + 1 \to N)} =  \frac{e^{-\beta \left( \Delta U + H_0 - A_0 - \mu^{\prime} \right)}}{N+1}.
\label{eq:ratio}
\end{equation} 
Within the simulation, we can easily determine the number of species present and the relevant potential and kinetic energies, and we externally set \(\mu^{\prime}\).
However, the absolute free energy \(A_{0}\) present in Eq.~\eqref{eq:ratio} is not known \emph{a priori} and is potentially difficult to calculate.
Instead, we would like to recast the equations in terms of a free energy difference, which is more computationally accessible.
In this spirit, we further divide the Monte Carlo entropy production \(\Sigma_{\mathrm{MC}} = \Sigma_{\mathrm{GCMC}} + \Sigma_{\mathrm{rxn}}\) into a term \(\Sigma_{\mathrm{GCMC}}\) associated with the details of the insertion or deletion procedure and a term \(\Sigma_{\mathrm{rxn}}\) associated with the fuel decomposition reaction FTC \(\ce{<=>}\) ETC + C.
By inserting Eq.~\eqref{eq:ratio} into Eq.~\eqref{eq:entprod}, summing over all GCMC moves, and separating the terms into terms that are unique to a specific GCMC move and those that are constant throughout the simulation, we arrive at two terms, the GCMC and reaction terms.
The GCMC term is given by
\begin{equation}
    \Sigma_{\mathrm{GCMC}} = -k_{\mathrm{B}} \sum_{i}^{N_{\mathrm{GCMC}}} \Delta_{i} \left[ \ln{ \left( N_{i}+1 \right) } +\beta \left( \Delta U_i + H_{i, 0}\right) \right].
\label{eq:gcmc_ent}
\end{equation}
The reaction term constitutes the entropy production due to the reaction,
\begin{equation}
\Sigma_{\mathrm{rxn}} = k_{\mathrm{B}} \beta \sum_{i}^{N_{\mathrm{GCMC}}} \Delta_{i} \left( \mu^{\prime} + A_{0} \right).
\end{equation}
Both the free energy \(A_{0}\) and the external chemical potential \(\mu^{\prime}\) are fixed, so this sum can be expressed in terms of the total number of insertions \(N_{i}^{+}\), deletions \(N_{i}^{-}\) and the free energy \(A_{0,i}\) of each species as
\begin{equation}
\Sigma_{\mathrm{rxn}} = k_{\mathrm{B}} \beta \Big[ (N^{+}_{\mathrm{FTC}}-N^{-}_{\mathrm{FTC}})(\mu^{\prime}_{\mathrm{FTC}} + A_{0,\mathrm{FTC}}) + (N^{+}_{\mathrm{ETC}}-N^{-}_{\mathrm{ETC}})(\mu^{\prime}_{\mathrm{ETC}} + A_{0,\mathrm{ETC}}) + (N^{+}_{\mathrm{C}}-N^{-}_{\mathrm{C}})(\mu^{\prime}_{\mathrm{C}} + A_{0,\mathrm{C}}) \Big].
\end{equation}
Since the system is in a steady state, we can use a species balance argument to simplify the right-hand side.
On average, the number of reactant molecules entering the system will be related to the number of product molecules leaving the system by the stoichiometry of the reaction taking place within the system.
These balances require that
\begin{align}
    \frac{\mathrm{d} }{\mathrm{d}t} \langle N_{\mathrm{FTC}} \rangle &= \lim_{t_{\mathrm{obs}} \to \infty } \frac{1}{t_{\mathrm{obs}}} \left( N^{+}_{\mathrm{FTC}}  -  N^{-}_{\mathrm{FTC}}  -  N_{\mathrm{rxn}} \right) = 0\\
   \frac{\mathrm{d} }{\mathrm{d}t} \langle N_{\mathrm{ETC}} \rangle &= \lim_{t_{\mathrm{obs}} \to \infty } \frac{1}{t_{\mathrm{obs}}} \left( N^{+}_{\mathrm{ETC}}  -  N^{-}_{\mathrm{ETC}}  +  N_{\mathrm{rxn}} \right) = 0\\
   \frac{\mathrm{d} }{\mathrm{d}t} \langle N_{\mathrm{C}} \rangle &= \lim_{t_{\mathrm{obs}} \to \infty } \frac{1}{t_{\mathrm{obs}}} \left( N^{+}_{\mathrm{C}}  -  N^{-}_{\mathrm{C}}  +  N_{\mathrm{rxn}} \right) = 0.
\end{align}
Here, the number of net reactions is given by the number of decomposition events (FTC \(\to\) ETC + C) less the number of recombination events (ETC + C \(\to\) FTC).
For a sufficiently well-established steady state (large \(t_{\mathrm{obs}}\)), we can relate the net flux of each species to the net number of reactions, \(N_{\mathrm{rxn}} = N_{\mathrm{FTC}}^{+} - N_{\mathrm{FTC}}^{-} =    N_{\mathrm{ETC}}^{-} - N_{\mathrm{ETC}}^{+} = N_{\mathrm{C}}^{-} - N_{\mathrm{C}}^{+} \).
The reaction entropy production can then be expressed succinctly as
\begin{equation}
\Sigma_{\mathrm{rxn}} = \frac{N_{\mathrm{rxn}}}{T} \left[ \mu^{\prime}_{\mathrm{FTC}}- \mu^{\prime}_{\mathrm{ETC}} - \mu^{\prime}_{\mathrm{C}} -\Delta A_{\mathrm{rxn},0} \right].
\label{eq:rxn_ent}
\end{equation}
We have introduced \(\Delta A_{\mathrm{rxn},0} = A_{0,\mathrm{ETC}} + A_{0,\mathrm{C}} - A_{0,\mathrm{FTC}}\) as the standard state free energy change of the reaction FTC \(\to\) ETC + C.
The problematic absolute free energies present in Eq.~\eqref{eq:ratio} have now been replaced by a free energy difference using a steady-state approximation.
Section~\ref{sec:free_en} details how this free energy difference is calculated.

We have thus broken down the total entropy production of the simulation into three terms: a contribution from the Langevin dynamics, a contribution from the GCMC insertion and deletion moves, and a contribution due to the reactions that take place,
\begin{equation}
%\Sigma_\mathrm{tot} = \Sigma_{\mathrm{Langevin}} + \Sigma_{\mathrm{GCMC}} + \Sigma_{\mathrm{rxn}}.
\Sigma = \Sigma_{\mathrm{Langevin}} + \Sigma_{\mathrm{GCMC}} + \Sigma_{\mathrm{rxn}}.
\label{eq:ent_tot}
\end{equation}
We calculate the Langevin contribution \(\Sigma_{\mathrm{Langevin}}\) using Eq.~\eqref{eq:langevin_ent}, the GCMC contribution \(\Sigma_{\mathrm{GCMC}}\) using Eq.~\eqref{eq:gcmc_ent}, and the reaction contribution \(\Sigma_{\mathrm{rxn}}\) using Eq.~\eqref{eq:rxn_ent}.
Each equation uses quantities that are easily accessible throughout the simulation, set externally, or can be calculated using straightforward methods.
\begin{figure}
\centering
\begin{tikzpicture}
    \node[anchor=south west,inner sep=0] (image) at (0,0) {
      \includegraphics[height=0.255\textwidth]{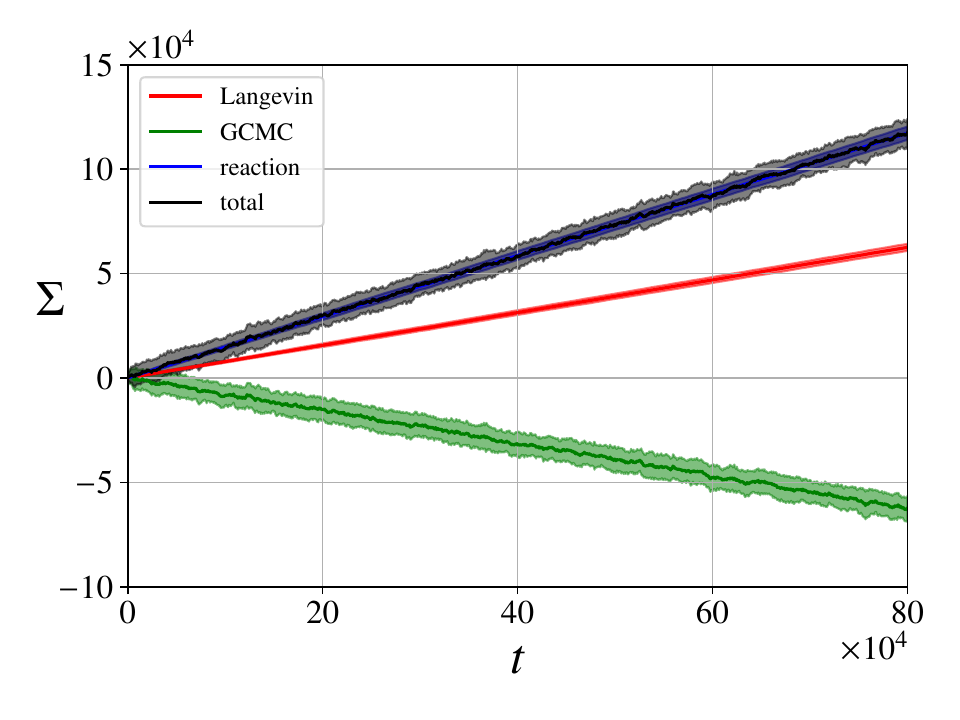}
    \includegraphics[height=0.26\textwidth]{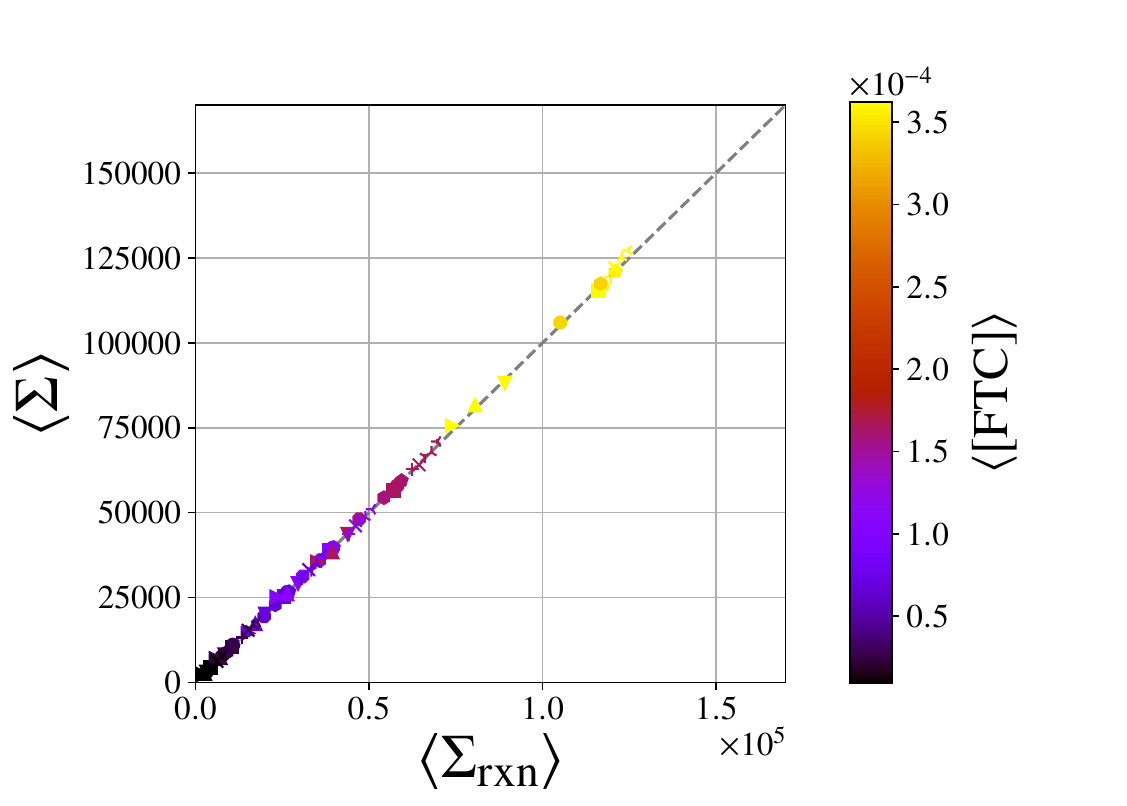}
    };
    \begin{scope}[x={(image.south east)},y={(image.north west)}]
        \node[] at (-0.02, 0.95) {(a)};
        \node[] at (0.5, 0.95) {(b)};
    \end{scope}
\end{tikzpicture}

\caption{(a) Total cumulative entropy production (black, Eq.~\eqref{eq:ent_tot}) and component cumulative entropy production for Langevin contributions (red, Eq.~\eqref{eq:langevin_ent}), GCMC contributions (green, Eq.~\eqref{eq:gcmc_ent}), and reaction contributions (blue, Eq.~\eqref{eq:rxn_ent}) over the course of a simulation upon reaching steady state.
The simulation uses \(\mu_{\mathrm{FTC}}^{\prime}=1.0\), \(\mu_{\mathrm{ETC}}^{\prime}=-10\), \(\mu_{\mathrm{C}}^{\prime}=-10\), and \(\gamma=0.5\) with a motor with \(N_{\mathrm{motif}}=2\).
Simulations are \(5 \times 10^{8}\) time steps long with a time step size of \(\Delta t = 0.002\) with the plot showing the last \(4 \times 10^8\) time steps after a steady state has been reached.
Solid lines represent the mean and shaded areas represent the standard deviation over 50 independent trajectories.
(b) The averaged total entropy production \(\langle \Sigma \rangle\) is equal to the averaged reaction entropy production \(\langle \Sigma_\mathrm{rxn} \rangle\),  \(\langle \Sigma \rangle \approx \langle \Sigma_\mathrm{rxn} \rangle\), verified for a family of catenane motors with varying numbers of binding and catalytic sites, friction, and backbone rigidity, denoted with different symbols (see Section~\ref{sec:deets} for details). The color bar shows the concentration of FTC molecules, which is controlled by the associated chemical potential.}
\label{fig:traj}
\end{figure}

Fig.~\ref{fig:traj}a shows the total cumulative entropy production and its components over the course of a motor simulation.
At steady state, the total and component entropy produced increase and decrease linearly.
As we can see, the mean entropy production rate is equal to \(\langle\Sigma_{\mathrm{rxn}}\rangle\), with \(\langle\Sigma_{\mathrm{GCMC}}\rangle\) and \(\langle\Sigma_{\mathrm{Langevin}}\rangle\) being equal in magnitude and opposite in sign:
\begin{equation}
    \langle\Sigma \rangle = \frac{\langle N_\mathrm{rxn}\rangle}{T}[\mu_\mathrm{FTC}-\mu_\mathrm{ETC}-\mu_\mathrm{C}],
    \label{eq:totentrisrxn}
\end{equation}
where \(\langle N_\mathrm{rxn}\rangle\) is the average net number of fuel decomposition reactions.
Physically, \(\Sigma_{\mathrm{rxn}}\) is the particle reservoir entropy production associated with having shifted FTC, ETC, and C from one reservoir to another via \(N_{\rm rxn}\) reactions.
The entropy produced in the particle reservoirs depends not only on how many net reactions occurred, but also on the energies of the species at the time they enter and exit the system.
The entropy production \(\Sigma_\mathrm{GCMC}\) measures the heat flows associated with addition and removal from the particle reservoirs.
Finally, \(\Sigma_\mathrm{Langevin}\) is the entropy production associated with the heat flow into a thermal reservoir for the Langevin timesteps during which no particles are added or removed to the system.
In the steady state, the energy flows into and out of the system must be balanced, so we must have \(-\langle\Sigma_\mathrm{GCMC}\rangle = \langle\Sigma_\mathrm{Langevin}\rangle\).
The stochastic simulations give direct access to \(\Sigma\), and its mean and variance can be explicitly computed.
If we only require \(\langle\Sigma \rangle\), Eq.~\eqref{eq:totentrisrxn} provides a shortcut for computing it by only counting reaction events.
Fig.~\ref{fig:traj}b further demonstrates \(\langle \Sigma \rangle \approx \langle \Sigma_\mathrm{rxn} \rangle\) for all simulated catenane motors with various designs and conditions.

\section{Calculations for the Minimal Model}
\label{sec:num_anal_minimal}
%\ggcom{[Change n to N variables. Describe Figure S3]}

In this section, we explain how we obtain the parameters for the minimal model from the particle simulations and from the thermodynamics of fuel decomposition. Using these parameters, we verify that the minimal model reproduces the results of the molecular simulation. 
This verification is particularly important because the minimal model operates on simplified dynamics that do not correspond to the actual physical mechanisms in play. Nonetheless, the minimal model describes the essential features for current generation in such systems and, owing to its simplicity, provides a straightforward physical interpretation for each parameter. We make use of this interpretation to link the minimal model and particle simulations.

We reiterate that \(\zeta\) is the combined rate of unbiased movement events; \(\lambda\), the combined rate of fuel transformation events; \(\eta\), the proportion of fuel transformation coupled to biased motion; and \(\Delta \mu\), the change in free energy associated with the decomposition of a single FTC species in solution.
Subsequently, we derive expressions for the mean and variance of the entropy production and current. 
Entropy is produced when the forward and backward rates of a given move are unequal.
This occurs for fuel decomposition both when futile and when coupled to the motion of the shuttling ring.
All states in the system are degenerate, and the shuttling ring's internal energy does not change.
Thus, all the free energy generated by fuel decomposition is released into the environment as heat.
The heat \(q_{k^+}\) generated along a path with forward rate \(k^+\) and reverse rate \(k^-\) is therefore
\begin{equation}
    \beta q_{k^+} = \ln\frac{k^+}{k^-}.
\end{equation}

We let \(N_0^+\), \(N_0^-\), \(N_\mathrm{fut}^+\), \(N_\mathrm{fut}^-\), \(N_\mathrm{bias}^+\), and \(N_\mathrm{bias}^-\) respectively be the number of forward and reverse moves within \(t_\mathrm{obs}\) associated with unbiased motion, futile fuel decomposition, and fuel decomposition coupled to biased motion.
These variables are all Poisson distributed with parameters equal to the rates listed in Fig.\ 2 and in the main text; for instance, \(N_0^+ \sim \mathrm{Poisson}(r_0t_\mathrm{tobs})\). 

Using these variables, we can write the current and the number of reactions as
%\ggcom{[Using \(\langle \cdot \rangle\) or \(\langle \cdot_\mathrm{S} \rangle\) or \(\langle \cdot \rangle_\mathrm{S}\)]}
\begin{align}
   % \Sigma &= (N_\mathrm{fut}^+ - N_\mathrm{fut}^-)\frac{q_\mathrm{fut}^+}{T} + (N_\mathrm{bias}^+ - N_\mathrm{bias}^-)\frac{q_\mathrm{bias}^+}{T}\\
    J &= (N_\mathrm{bias}^+ - N_\mathrm{bias}^- + N_0^+ - N_0^-)l\\
    N_\mathrm{rxn} &= N^+_{\rm bias} - N^-_{\rm bias} + N_{\rm fut}^+ - N_{\rm fut}^-
\end{align}
Taking means and variances of these two equations yields
\begin{align}
 \langle J\rangle &= (r_\mathrm{bias}^+ - r_\mathrm{bias}^- + r_0 - r_0)t_\mathrm{obs}l\\
    \text{var}(J) &= (r_\mathrm{bias}^+ + r_\mathrm{bias}^- + r_0 + r_0)t_\mathrm{obs}l^2\\
    \langle N_{\rm rxn }\rangle &= (r_\mathrm{bias}^+ - r_\mathrm{bias}^-+ r_\mathrm{fut}^+ - r_\mathrm{fut}^-)t_\mathrm{obs}\\
    \text{var}(N_{\rm rxn }) &= (r_\mathrm{bias}^+ + r_\mathrm{bias}^-+ r_\mathrm{fut}^+ + r_\mathrm{fut}^-)t_\mathrm{obs}
    %\langle\Sigma\rangle &= (r_\mathrm{fut}^+ - r_\mathrm{fut}^-)t_\mathrm{obs}\frac{q_\mathrm{fut}^+}{T} + (r_\mathrm{bias}^+ - r_\mathrm{bias}^-)t_\mathrm{obs}\frac{q_\mathrm{bias}^+}{T}\\
    %\text{var}(\Sigma) &= (r_\mathrm{fut}^+ + r_\mathrm{fut}^-)t_\mathrm{obs}\left(\frac{q_\mathrm{fut}^+}{T}\right)^2 + (r_\mathrm{bias}^+ + r_\mathrm{bias}^-)t_\mathrm{obs}\left(\frac{q_\mathrm{bias}^+}{T}\right)^2\\
\end{align}
by making use of the fact that \(\langle N_\mathrm{fut}^+\rangle = \text{var}(N_\mathrm{fut}^+) = r_\mathrm{fut}^+t_\mathrm{obs}\) for \(N_\mathrm{fut}^+ \sim \mathrm{Poisson}(r_\mathrm{fut}^+t_\mathrm{obs})\), and so on.

Next, we must identify expressions for the forward and reverse rates in each case. We begin by evaluating \(r_\mathrm{bias}^+\) and \(r_\mathrm{bias}^-\) given the constraints that
\begin{equation}
    \frac{r_\mathrm{bias}^+}{r_\mathrm{bias}^-} = \exp(\beta\Delta\mu),
\end{equation}
because fuel decomposition produces a driving force \(\Delta\mu\), and, by construction, 
\begin{equation}
    r_\mathrm{bias}^+ + r_\mathrm{bias}^- = \eta\lambda.
\end{equation}
Solving this system of two equations and two unknowns recovers
\begin{align}
    r_\mathrm{bias}^+ &= \frac{\eta\lambda}{1+e^{-\beta\Delta\mu}}\\
    r_\mathrm{bias}^- &= \frac{\eta\lambda e^{-\beta\Delta\mu}}{1+e^{-\beta\Delta\mu}}.
\end{align}

Finally, we verify that the results of our model are consistent with those from numerical simulation. In the particle simulation, during an observation time \(t_\mathrm{obs}\), let the shuttling ring make \(N^+\) moves forward in the favored direction and \(N^-\) moves backward, and let \(N_\mathrm{fuel}\) fuel transformation events take place.  The minimal model comprises the superposition of a perfectly biased and a perfectly unbiased random walker. Since \(\Delta \mu \gg 0\) and the amount of fuel reconstitution events is negligible, we attribute \(N^+ - N^-\) moves to the biased random walker and \(2N^-\) moves to the unbiased random walker. Then, following the physical interpretation of these parameters, we compute the effective parameters from the simulation with \(\zeta_\mathrm{eff} = 2\langle N^- \rangle /t_\mathrm{obs}\), \(\lambda_\mathrm{eff}= \langle N_\mathrm{fuel} \rangle/t_\mathrm{obs}\), and \(\eta_\mathrm{eff} = (\langle N^+\rangle -\langle N^- \rangle )/N_\mathrm{fuel}\). \(\Delta \mu\) follows from the thermodynamics of fuel decomposition; we treat this parameter in detail in Section~\ref{sec:free_en}.

\begin{figure}
\centering
\begin{tikzpicture}
    \node[anchor=south west,inner sep=0] (image) at (0,0) {
      \includegraphics[height=0.25\textwidth]{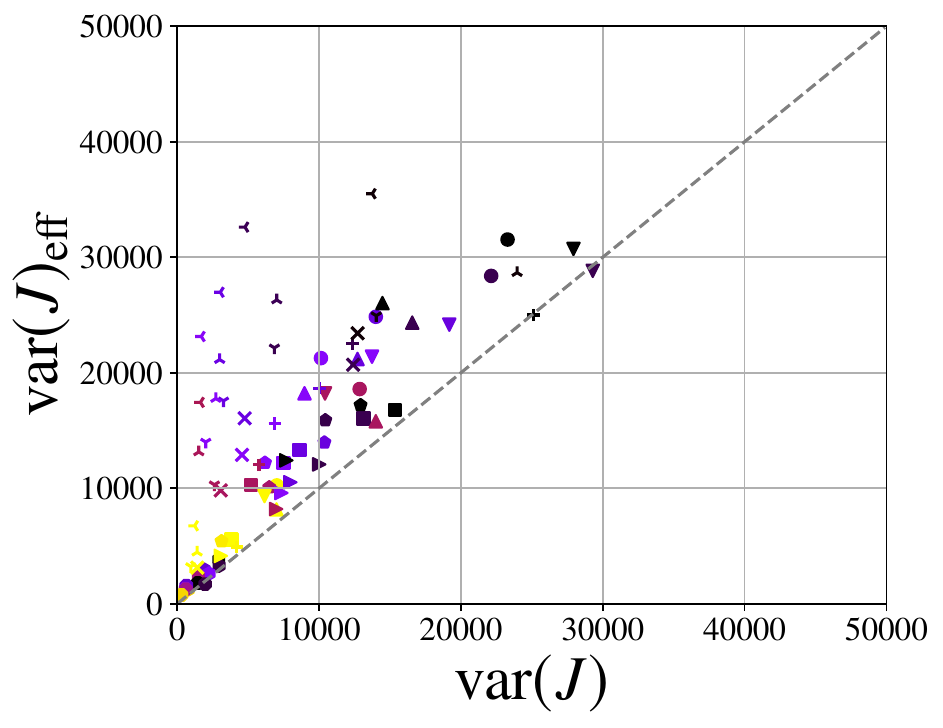}
    \hspace{0.1in}
    \includegraphics[height=0.277\textwidth]{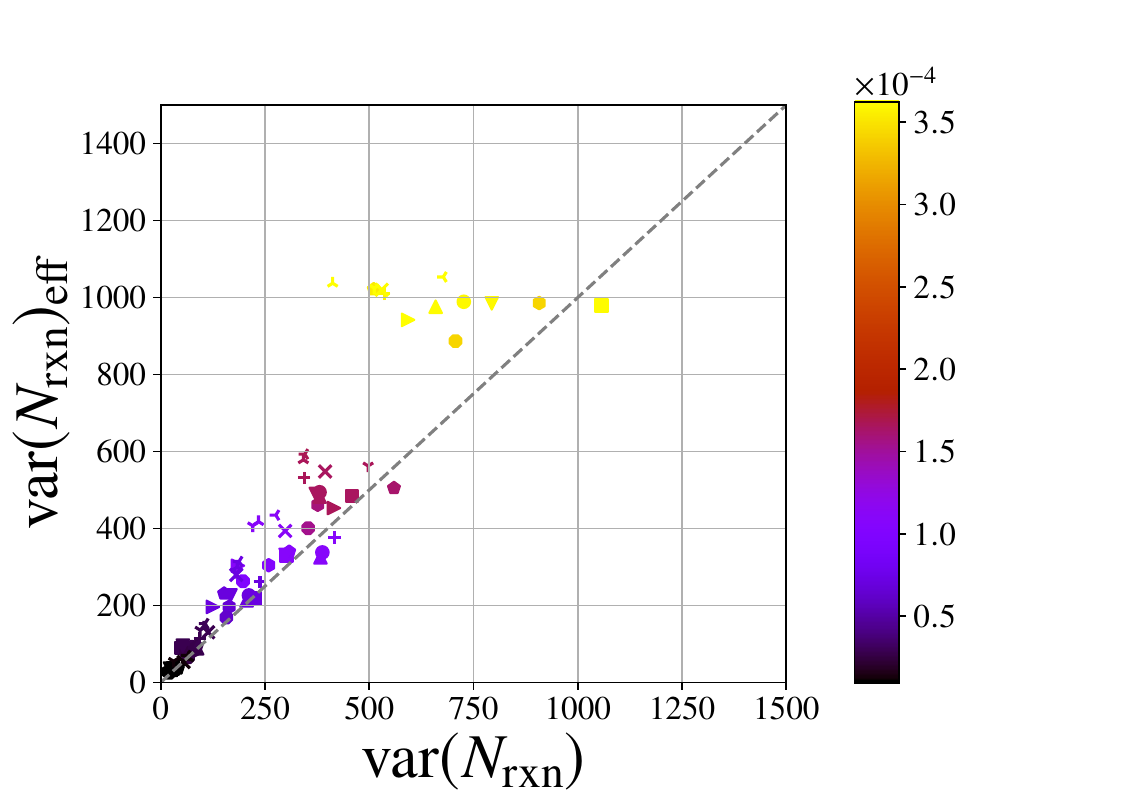}
    };
    \begin{scope}[x={(image.south east)},y={(image.north west)}]
        \node[] at (0, 0.95) {(a)};
        \node[] at (0.5, 0.95) {(b)};
    \end{scope}
\end{tikzpicture}
\caption{Comparison of the computed variances \(\langle J \rangle_\mathrm{eff}\) (\emph{a}) and \(\langle N_\mathrm{rxn} \rangle_\mathrm{eff}\) (\emph{b}) from the minimal model using Eqs.~(9) and (6) against the simulated variances \(\langle J \rangle\) and \(\langle N_\mathrm{rxn} \rangle\). The symbols represent a family of catenane motors with varying numbers of binding and catalytic sites, friction, and backbone rigidity, denoted with different symbols (see Section~\ref{sec:deets} for details). The color bar shows the concentration of FTC molecules, which is controlled by the associated chemical potential.} 
\label{fig:S3}
\end{figure}

The simulated averages \(\langle J \rangle\) and \(\langle N_{\rm rxn} \rangle\) are approximately equal to \(\langle J \rangle_\mathrm{eff}\) and \(\langle N_{\rm rxn} \rangle_\mathrm{eff}\) computed in the model, since \(\langle J \rangle_\mathrm{eff} \approx \eta_\mathrm{eff} \lambda_\mathrm{eff}t_{\rm obs}l  = (\langle N^\mathrm{+} \rangle - \langle N^\mathrm{-} \rangle)l=\langle J \rangle \) and  \(\langle N_{\rm rxn} \rangle_\mathrm{eff} \approx  \lambda_{\rm eff} t_{\rm obs}= \langle N_\mathrm{fuel} \rangle \). However, \(\mathrm{var}(J)_\mathrm{eff}\) and \(\mathrm{var}(N_{\rm rxn})_\mathrm{eff}\) are independently computed from the minimal model using Eqs.~(9) and (6) with effective parameters fitted from the simulations. Fig.~\ref{fig:S3} demonstrates quantitative agreement between the computed and simulated variances.

\section{Free Energy and Entropy of the Decomposition Reaction}
\label{sec:free_en}
We consider here the thermodynamic properties of fuel decomposition, particularly regarding the calculation of \(\Delta\mu\), the change in free energy associated with the decomposition of a single FTC species in solution. 
We can write
\begin{equation}
    \Delta\mu = \mu_\mathrm{FTC}-\mu_\mathrm{ETC}-\mu_\mathrm{C}
\end{equation}
following Eq.~\eqref{eq:totentrisrxn}, but it turns out to be more convenient operationally to follow the alternate decomposition of Eq.~\eqref{eq:rxn_ent},
\begin{equation}
    \Delta\mu = \mu^{\prime}_{\mathrm{FTC}}- \mu^{\prime}_{\mathrm{ETC}} - \mu^{\prime}_{\mathrm{C}} - \Delta A_{\mathrm{rxn},0},
\end{equation}
where \(\mu'_i = \mu_i - A_{0,i}\) is the external or relative chemical potential of species \(i\), \(A_{0,i}\) is the free energy of a single species \(i\) in the simulation volume and temperature, and \(\Delta A_{\mathrm{rxn},0} = A_{0,\mathrm{ETC}} + A_{0,\mathrm{C}} - A_{0,\mathrm{FTC}}\) is the standard free energy of fuel decomposition.
In the simulation, the various \(\mu'\) are set externally, and \(\Delta A_{\mathrm{rxn},0}\) is a thermodynamic constant that we now evaluate.

The reaction \(\mathrm{FTC \ce{<=>} ETC + C}\) has the associated change in entropy
\begin{equation}
\Delta S_{\mathrm{rxn},0}  = \frac{\Delta U_{\mathrm{rxn},0}  - \Delta A_{\mathrm{rxn},0}}{T},
\label{eq:DeltaS}
\end{equation}
where \(\Delta U_{\mathrm{rxn},0}\) is the energy difference between a single FTC and single ETC and C in isolation, and \(T\) is the temperature. 
We sampled configurations of \(\mathrm{FTC}\), \(\mathrm{ETC}\) and \(\mathrm{C}\) under the canonical ensembles to calculate \(\Delta U_\mathrm{rxn} = \langle U_\mathrm{ETC} \rangle + \langle U_\mathrm{C} \rangle  -  \langle U_\mathrm{FTC} \rangle\). The corresponding change in the Helmholtz free energy change is given by
\begin{equation}
    \Delta A_{\mathrm{rxn},0} = k_\mathrm{B}T\ln \frac{Z_\mathrm{ETC} Z_\mathrm{C} }{Z_\mathrm{FTC}}, \label{eq:dArxn}
\end{equation}
where \(Z\) values are the partition functions of each species. 
Eq.~\eqref{eq:dArxn} assumes that the products ETC and C are completely non-interacting in this finite-box simulation. 
In practice, we set the threshold for non-interactivity at a separation distance \(d > 4\) between the center of mass of the ETC and C species. 
We further set the threshold for the bound species FTC at a separation distance \(d < 0.25\), for which the C species is entirely trapped within the ETC species, and in between is the transition region. Thus,
\begin{equation}
\Delta A_{\mathrm{rxn},0} = -k_\mathrm{B}T\ln \frac{Z(d > 4)}{Z(d < 0.25)} = -k_\mathrm{B}T \ln \frac{P(d > 4)}{P(d < 0.25)}
\label{eq:DeltaA}
\end{equation}
where \(Z(d > 4)\) and \(Z(d < 0.25)\) are the canonical partition functions of the product and reactant states, and \(P(d > 4)\) and \(P(d < 0.25)\) are the associated probabilities for \(d\).

We used the multistate Bennett acceptance ratio (MBAR) estimator to compute this free energy difference. 
MBAR combines data from multiple simulations under a series of biasing potentials to reweight samples for the unbiased potential~\cite{shirts2008statistically}. 
Then, the probabilities of the reactant and product states are calculated based on these weights. 
We designated sampling windows that are evenly located along the separation distance \(d\) by applying extra harmonic potentials of the form
\begin{equation}
    \Delta V = \frac{1}{2} k (d-d_c)^2,
\end{equation}
where \(k\) is the strength of the harmonic potential and \(d_c\) is the center of the potential. 

\begin{figure}
\centering
\includegraphics[width=0.45\textwidth]{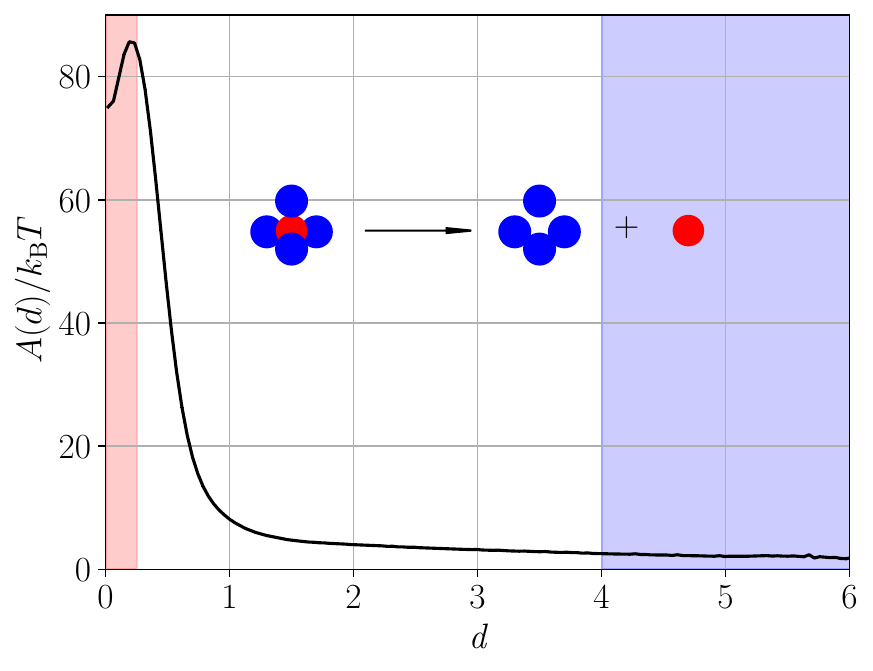}
\caption{
Free energy profile along the separation distance \(d\).
The red and blue regions respectively represent the reactant and product states.
}
\label{fig:FE}
\end{figure}

We used Monte Carlo (MC) simulations to sample \(d\) under each biasing potential, with the box size fixed at $34\times 34\times 34$ (as in the non-equilibrium simulation), \(T = 0.5\) and with periodic boundary conditions. 
Two sets of \(k\) and \(d_c\) were chosen to perform simulations: \(k=0, d_c = 0\) (unbiased sampling) and \(k = 1000, d_c = 0, 0.05, \dotsc, 2\) (biased sampling). 
We chose \(1\times 10^5\) samples from the MC simulations with \(2\times 10^8\) steps. 
The free energy profile along the separation distance \(d\) is shown in Fig.~\ref{fig:FE}, yielding \(\Delta U_{\mathrm{rxn}, 0} = - 63.2 k_\mathrm{B}T \), \(\Delta A_{\mathrm{rxn}, 0} = - 76.4 k_\mathrm{B}T \), and \(\Delta S_{\mathrm{rxn}, 0} = 13.2 k_\mathrm{B}\).

\section{Bulk Contributions to the TUR Bound}
\label{sec:additional}
In this section, we consider whether the deviation from the TUR bound might be explained by futile fuel transformation in the bulk, corresponding to futile cycles in the minimal model.
As discussed in Section~\ref{sec:derive} and shown in Fig.~\ref{fig:traj}, the total entropy production rate is roughly equal to the entropy production rate due to reactions alone.
Since the decomposition of FTC can occur both in the bulk and at the motor's catalytic sites, it is natural to separate the local entropy production caused by the motor from background entropy production in the bulk.
By repeating our analysis on the background-subtracted data, we demonstrate that, while the local entropy production is necessarily less than the entropy production of the whole system, the motor remains far from the TUR limit across a range of conditions and configurations.

\begin{figure*}
\centering
\includegraphics[width=0.45\textwidth]{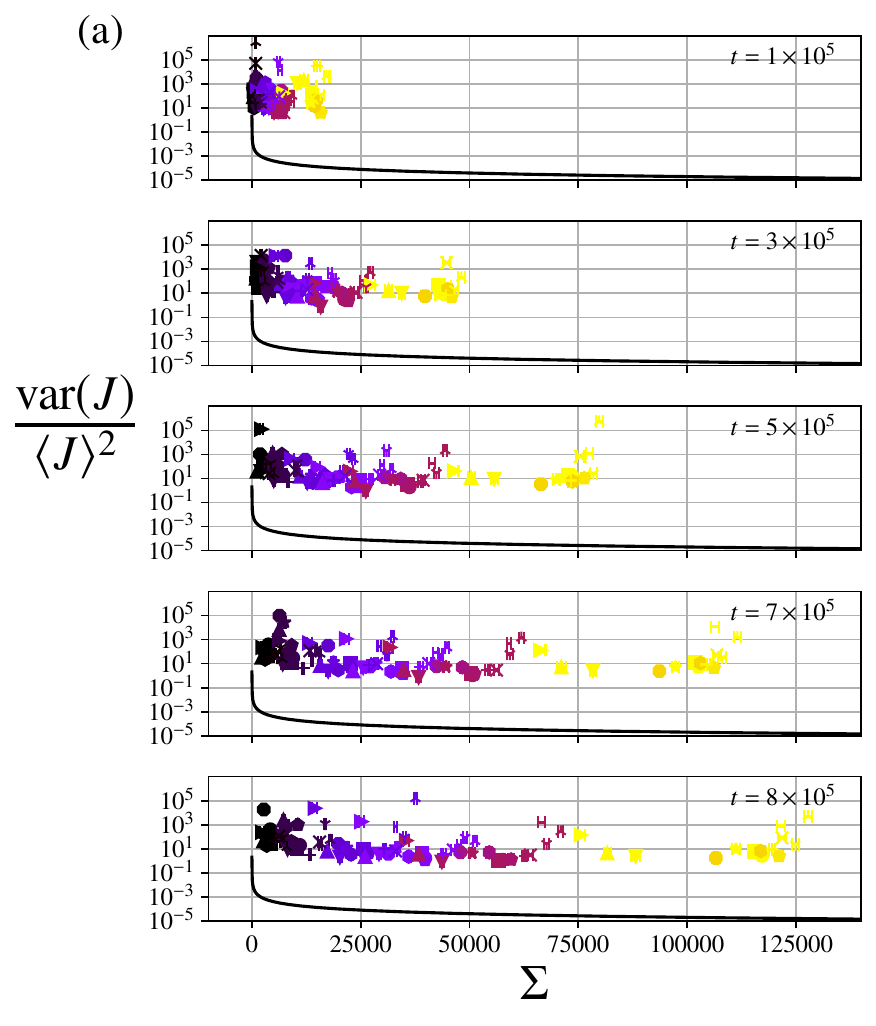} 
\includegraphics[width=0.45\textwidth]{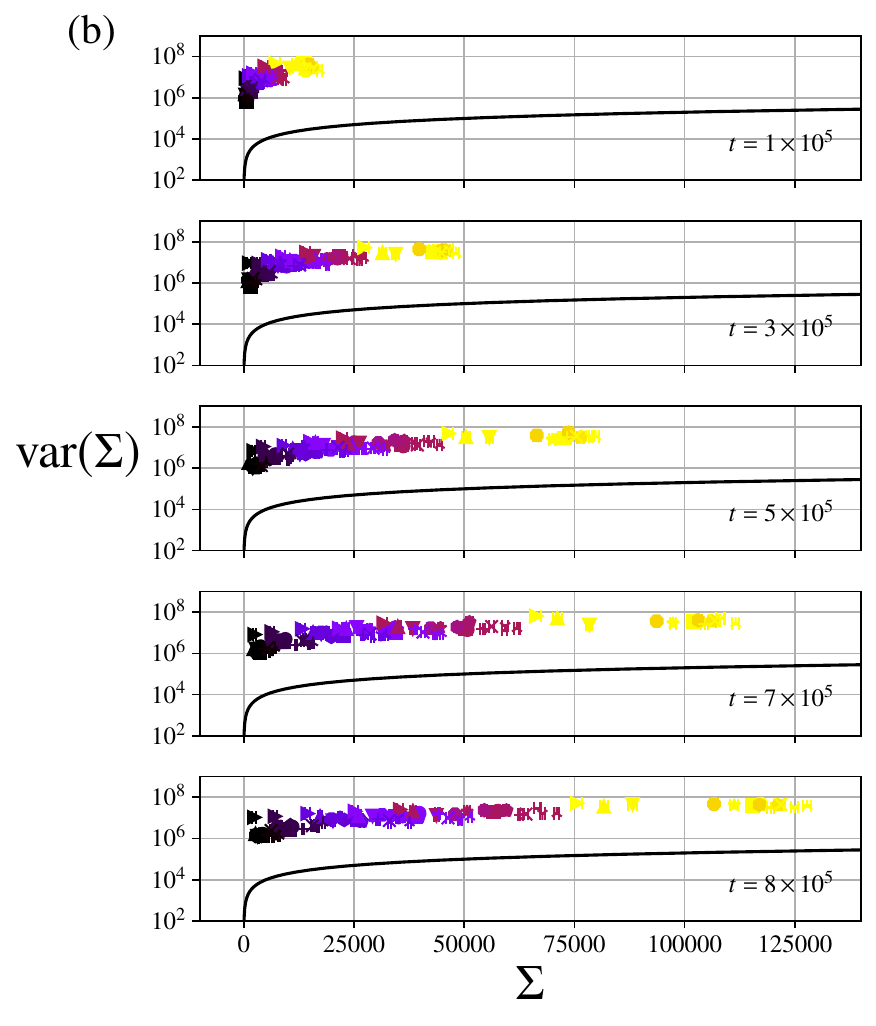}  \\
\hspace{0.15cm}\includegraphics[width=0.469\textwidth]{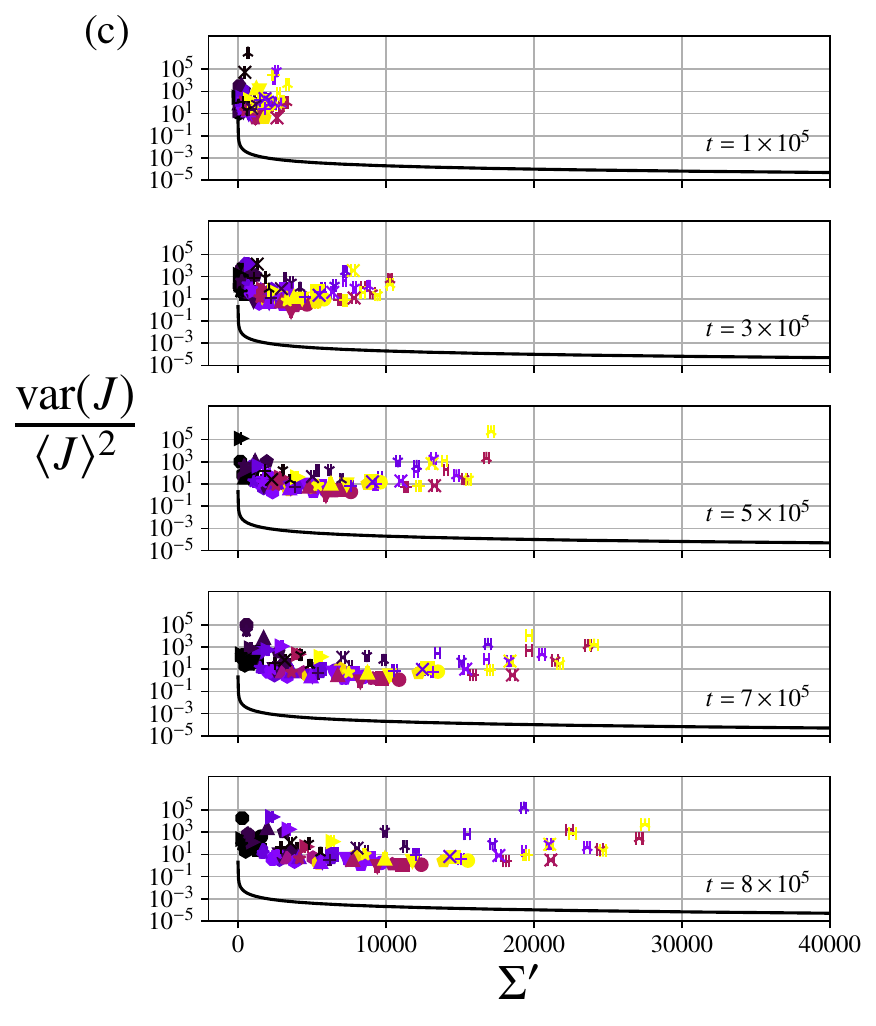}  
\hspace{-0.35cm}\includegraphics[width=0.469\textwidth]{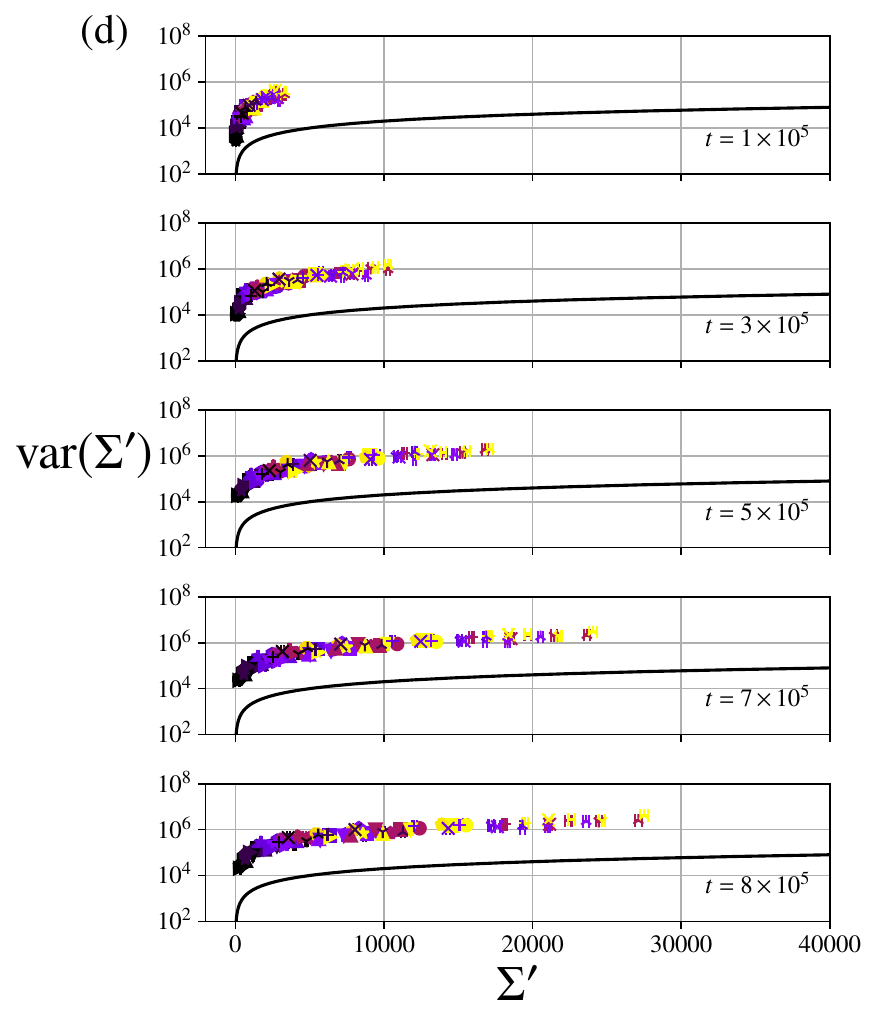} \\ 
\hspace{1.5cm}\includegraphics[width=0.6\textwidth]{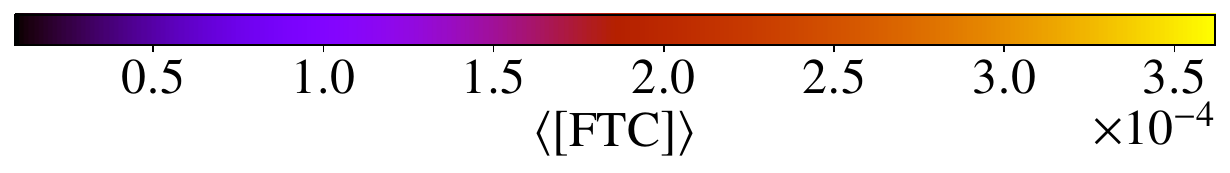}  
\caption{(\emph{a}) Motor precision compared to the current TUR using total entropy production, (\emph{b}) entropy variance compared to entropy TUR using total entropy production, (\emph{c}) motor precision compared to the current TUR using local entropy production, and (\emph{d}) entropy variance compared to entropy TUR using local entropy production.
Data is shown at intermediate times along a \(t_{\mathrm{obs}}=10^{6}\) trajectory chronologically from top to bottom.
Symbols and colors agree with those of Fig.\ 1 and are described in Section~\ref{sec:deets}.
Means and variances were collected over 50 independent simulations at each data point.
Error bars in the average entropy production are the standard error.}
\label{fig:times}
\end{figure*}

We define the localized motor entropy production in terms of the net number of catalyzed reactions \(N_{\mathrm{cat,rxn}}\):
\begin{equation}
\Sigma^{\prime} = \frac{N_{\mathrm{cat,rxn}}}{T} \left[ \mu^{\prime}_{\mathrm{FTC}}- \mu^{\prime}_{\mathrm{ETC}} - \mu^{\prime}_{\mathrm{C}} -\Delta A_{\mathrm{rxn},0} \right].
\label{eq:local}
\end{equation}
\begin{figure}
\centering
\includegraphics[width=0.45\textwidth]{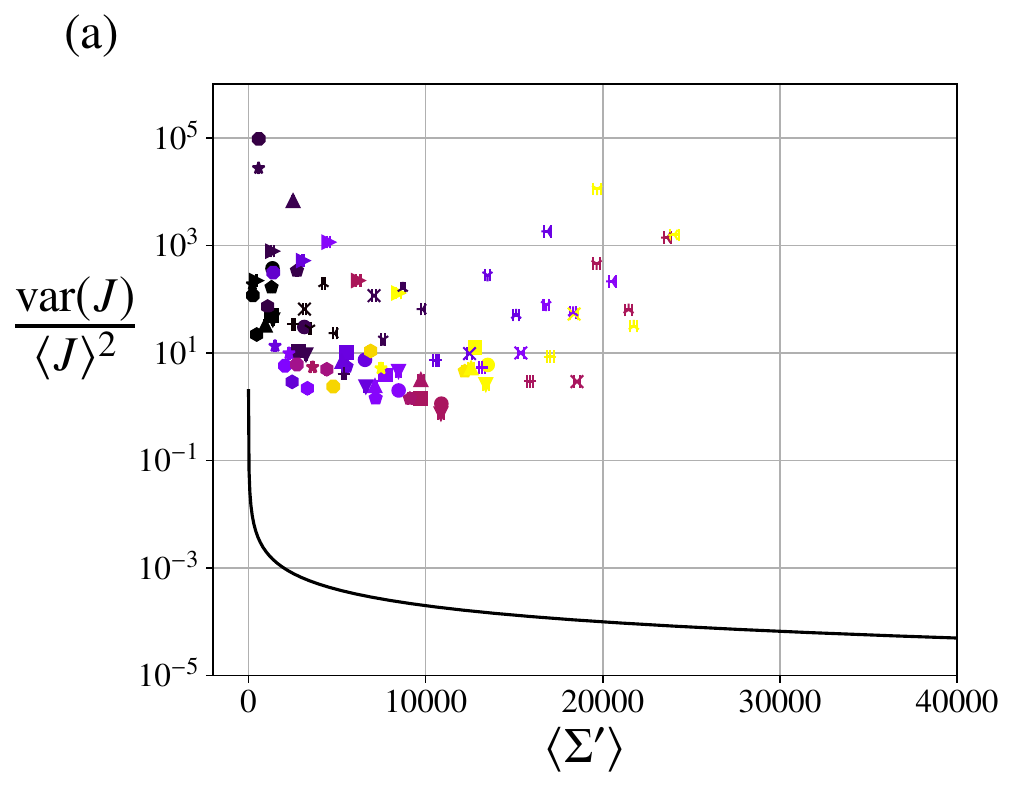}\\
\includegraphics[width=0.45\textwidth]{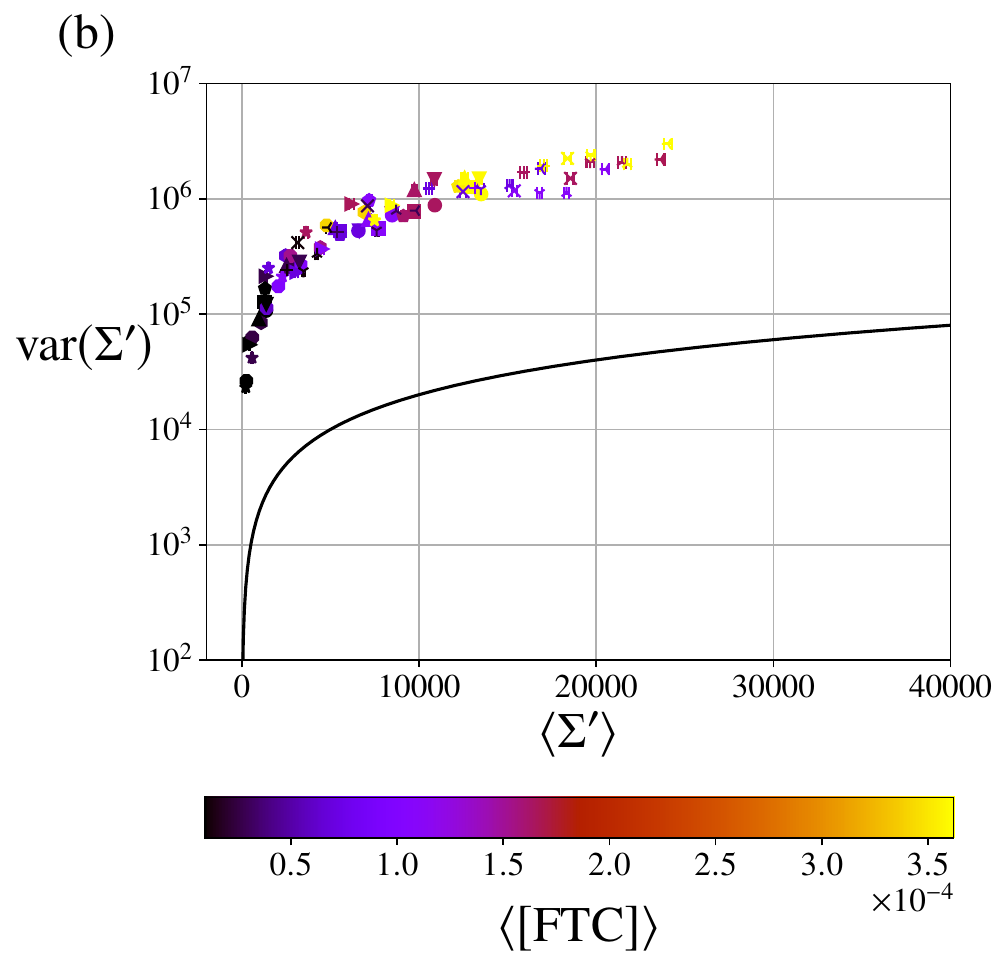}
\caption{
Simulated motor precision (\emph{a}) and produced entropy (\emph{b}) compared to the general TUR (Eq.\ (1)) and entropy TUR (Eq.\ (2)), respectively, using the local entropy production of Eq.~\eqref{eq:local} instead of the total entropy production.
Symbols and colors agree with those of Fig.\ 1 and are described in Section~\ref{sec:deets}.
Means and variances were collected over 50 independent simulations at each data point.
Error bars in the average entropy production represent standard error.
}
\label{fig:sim_cat_tur}
\end{figure}
A reaction is considered catalyzed when it occurs within 2 distance units of the middle particle of a catalytic site.
Otherwise, the reaction is considered to be in the bulk.
Following these definitions, Fig.~\ref{fig:sim_cat_tur} shows the motor's precision compared to the TUR bound when using the local entropy production in place of the total entropy production, and Fig.~\ref{fig:times} shows the simulated motor precision at intermediate time points for the current TUR and the entropy TUR using both the total and local entropy production.

\end{document}